\documentclass[%
 reprint,
superscriptaddress,
 amsmath,amssymb,
 aps,
]{revtex4-2}

\usepackage{xcolor}
\usepackage{graphicx}% Include figure files
\usepackage{dcolumn}% Align table columns on decimal point
\usepackage{bm}% bold math
\begin{document}

\preprint{APS/123-QED}

\title{Superconducting dome and structural changes in LaRu$_3$Si$_2$ under pressure}% Force line breaks with \\

\author{Z. Li}
\email{Contact author: zqli@iastate.edu}
\affiliation{%
 Ames National Laboratory, US Department of Energy, Iowa State University, Ames, Iowa 50011, USA \\
}%
\affiliation{%
 Department of Physics and Astronomy, Iowa State University, Ames, Iowa. 50011, USA\\
}%

\author{S. Huyan}%
\affiliation{%
 Ames National Laboratory, US Department of Energy, Iowa State University, Ames, Iowa 50011, USA \\
}%
\affiliation{%
 Department of Physics and Astronomy, Iowa State University, Ames, Iowa. 50011, USA\\
}%

\author{E. C. Thompson}%
\affiliation{%
Department of Physics, University of Alabama at Birmingham, Birmingham, AL 35294, USA\\
}%

\affiliation{%
Department of Earth and Environmental Systems, The University of the South, Sewanee, Tennessee 37383, USA\\
}%

\author{T. J. Slade}%
\affiliation{%
 Ames National Laboratory, US Department of Energy, Iowa State University, Ames, Iowa 50011, USA \\
}%

\author{D. Zhang}%
\affiliation{%
Center for Advanced Radiation Sources, The University of Chicago, Chicago, Illinois 60637,USA\\
}%

\author{Y. J. Ryu}%
\affiliation{%
Center for Advanced Radiation Sources, The University of Chicago, Chicago, Illinois 60637,USA\\
}%

\author{W. Bi}%
\affiliation{%
Department of Physics, University of Alabama at Birmingham, Birmingham, AL 35294, USA\\
}%

\author{S. L. Bud'ko}%
\affiliation{%
 Ames National Laboratory, US Department of Energy, Iowa State University, Ames, Iowa 50011, USA \\
}%
\affiliation{%
 Department of Physics and Astronomy, Iowa State University, Ames, Iowa. 50011, USA\\
}%

\author{P. C. Canfield}%
\email{Contact author: canfield@ameslab.gov}
\affiliation{%
 Ames National Laboratory, US Department of Energy, Iowa State University, Ames, Iowa 50011, USA \\
}%
\affiliation{%
 Department of Physics and Astronomy, Iowa State University, Ames, Iowa. 50011, USA\\
}%

\date{\today}

\begin{abstract}
LaRu$_3$Si$_2$ is of current research interest as a kagome metal with a superconducting transition temperature, $T_c\sim$7 K and higher-temperature charge density wave orders. Here we report on electrical transport and x-ray diffraction measurements on LaRu$_3$Si$_2$ under pressure up to 65 GPa and 35 GPa, respectively. The superconducting transition temperature $T_c$ first gets slightly enhanced and reaches a maximum $\sim$8.7 K at $\sim$8.5 GPa. With further applied pressure, $T_c$ is initially gradually suppressed, then more rapidly suppressed, followed by gradual suppression, revealing a superconducting dome. Two possible pressure-induced structural phase transitions are also observed at room temperature, from the original hexagonal phase to another hexagonal structure above $\sim$11.5 GPa, and further to a structure with lower symmetry above $\sim$23.5 GPa.  These transition pressures roughly correlate with features found in our pressure dependent transport data.

\end{abstract}

\maketitle

\section{Introduction}
Superconductivity in hexagonal LaRu$_3$Si$_2$ with a transition temperature $T_c$, $\sim$7 K, was discovered more than 40 years ago \cite{BARZ19801489,vandenberg1980crystal}. Initial studies of superconductivity in this compound were performed via La substitution by other (magnetic or strongly correlated) rare earths \cite{god84a,god87a,esc94a}. It should be noted that the observed $T_c$ suppression with heavy rare earth substitution (Gd, Tm) does not scale with the de Gennes factor \cite{god87a,esc94a}.  More recent studies of chemical doping in  LaRu$_3$Si$_2$ \cite{lib12a,lib16a,RhIr_Doping_2023} show that Y, Lu and Ce doping on the La site as well as Cr, Co, Ni, Cu, Ir, and Rh doping on the Ru site suppress $T_c$ very slowly, whereas Fe substitution for Ru causes fast $T_c$ suppression. That can be described by Abrikosov - Gor'kov theory since Fe ions presented as strong magnetic scattering centers \cite{abr60a,lib16a}.

In $^{139}$La-NQR measurements \cite{kis04a,kis04b} a clear coherence peak in the spin relaxation rate, $1/T_1$,  just below superconducting transition temperature, $T_c$, was observed. It was followed, on further cooling, by the exponential behavior of $1/T_1$. These results identified  LaRu$_3$Si$_2$ as a conventional strong-coupled BCS superconductor.  The energy gap was estimated as $2\Delta (0) = 5.50 k_B T_c$. The somewhat high $T_c$ was presumed to be due to the high density of states at the Fermi level combined with strong electron - phonon interaction.

This point of view was reexamined recently \cite{mie21a}, and based on muon spin rotation ($\mu$SR) experiments and first-principles calculations, other factors were suggested to explain the enhanced $T_c$, such as correlation effects from the kagome flat band or the Van Hove point on the kagome lattice. Further $\mu$SR measurements on La(Ru$_{1-x}$Fe$_x$)$_3$Si$_2$ \cite{mie24a} proposed that the superconductivity is unconventional and  that the Fe doping introduces nodes in the superconducting gap structure. Altogether these studies point to interesting questions related to kagome superconductivity and balance between competing orders associated with it \cite{nun20a,rom22a,aid24a}. 

Pressure is known as a \textquoteleft clean\textquoteright alternative to chemical substitution---a way to induce controlled perturbation in materials. A large body of work has studied the effects of pressure in kagome superconductors \cite{zho24a}. For LaRu$_3$Si$_2$, though, we are aware only of $\mu$SR studies up to just under 2 GPa \cite{mie21a,gug23a}. The main conclusion was that both $T_c$ and the superfluid density $\lambda^{-2}_{eff}$ of LaRu$_3$Si$_2$ are relatively insensitive to hydrostatic pressure in this range. Similarly, pressure-independent $T_c$ and superfluid density were observed in the $\mu$SR measurements on  La(Ru$_{0.98}$Fe$_{0.02}$)$_3$Si$_2$ \cite{mie24a} up to 1.94 GPa.

Being puzzled by the apparent robust behavior of $T_c$ in the $\sim$2 GPa pressure range, and keeping in mind recently discovered charge order at ambient pressure in La(Ru$_{1-x}$Fe$_{x}$)$_3$Si$_2$ \cite{plo24a}, in this work we study superconductivity in LaRu$_3$Si$_2$ by measuring electrical transport in a diamond anvil cell (DAC) at pressures up to 65.5 GPa, as well as its room-temperature structure by x-ray diffraction up to 35 GPa, thus exceeding the literature pressure range by more than an order of magnitude. We find that $T_c$ is actually resolvably pressure sensitive, reaching a peak near 8.5 GPa at 8.7 K and then dropping to 2 K or below at $>$40 GPa.  In addition, we can correlate the structural changes with features in the $T_c(P)$ diagram as well as features in the transport data.

\section{Experimental Details}

Polycrystalline LaRu$_3$Si$_2$ was prepared via arc-melting elemental La (ingot, Material Preparation Center - Ames National Laboratory 99.99\%), Ru (lump, Alfa Aesar 99.95\% (metal basis)), and Si (lump, Material Preparation Center - Ames National Laboratory 99.999\%) on a water-cooled copper hearth using Zr pellets as an oxygen getter. In order to suppress the second phase LaRu$_2$Si$_2$, excess Ru was added to reach the starting composition LaRu$_{3.15}$Si$_2$ \cite{BARZ19801489}. The melted button was flipped and thoroughly remelted five times to ensure a homogeneous melt. The total mass of the pellet was 1.86 g after the final melting, losing less than 1\% mass than before arc-melting. Part of the arc-melted button was then broken manually into pieces for further measurements.

The ambient pressure crystal structure was examined  by powder x-ray diffraction (PXRD) using a Rigaku MiniFlex \MakeUppercase{\romannumeral 2} powder diffractometer with Cu K$\alpha$ radiation ($\lambda$ = 1.5406 \r A). The pieces from the arc-melted button were crushed and ground to fine powder, and dispersed evenly on a single-crystal Si zero-background holder, with the aid of a small amount of vacuum grease. Intensities were collected for 2$\theta$ ranging from 15$^{\circ}$ to 100$^{\circ}$, in a step size of 0.015$^{\circ}$, counting for 4 s at each angle. Rietveld refinement was preformed on as-collected PXRD data using the GSAS \MakeUppercase{\romannumeral 2} software package \cite{Toby:aj5212_GSAS}.

Compositional analysis and elemental mapping were carried out using the JEOL NeoScope JCM-7000 Benchtop scanning electron microscope (SEM) on a polished sample piece, under an accelerating voltage of 15 kV. Elemental mapping was performed at 2500 magnification, with the resolution of 768$\times$1024 pixels. Quantitative analysis of the x-ray spectrum was done with the built-in software SMILE VIEW Map with factory standards. 

Magnetization measurements were done in a Quantum Design magnetic property measurement system (MPMS) superconducting quantum interference device (SQUID) magnetometer (operated in the temperature range from 1.8 K to 300 K and magnetic field from -70 kOe to 70 kOe). An irregularly shaped sample was measured under both zero field cooling (ZFC) an field cooling (FC) protocols. This sample was placed in between two uniform plastic straws.

High pressure electrical transport measurements were done in a DAC \cite{noauthor_23mm_nodate}, with 350 $\mu$m culet size standard cut type Ia diamonds. A piece of the broken arc-melted sample was polished into a 20 $\mu$m thick flake and cut into a 60 $\mu$m $\times$ 60 $\mu$m size plate, and loaded into the DAC. The sample was loaded together with two $<$10 $\mu$m diameter ruby spheres into an apertured stainless-steel gasket covered by cubic BN. Platinum foil was used to make electrical contacts on a sample flake to perform van der Pauw method resistance measurement. Nujol mineral oil was used as a pressure-transmitting medium (PTM). Two samples (S\#1 and S\#2) from the same batch were measured under the same high-pressure configuration and protocols.

\begin{figure}[bp]
\includegraphics[width=8cm]{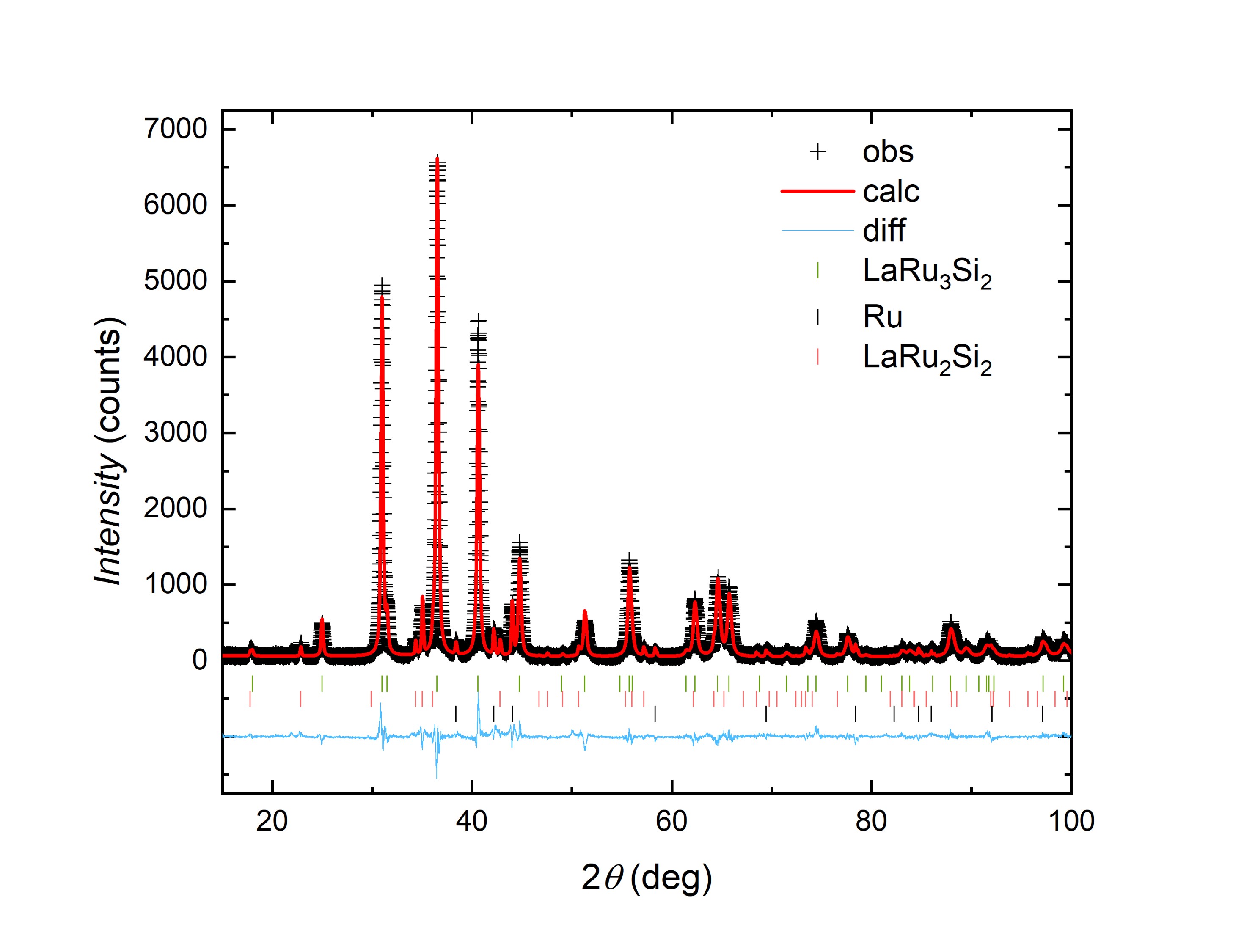}
\caption{\label{fig:PXRD} Ambient pressure x-ray diffraction pattern of LaRu$_3$Si$_2$; red line is the calculated fitting curve. All main diffraction peaks can be indexed well by a hexagonal structure with $a=5.676$\r A, $c = 3.558$\r A with Ru and LaRu$_2$Si$_2$ as impurity phases. 
}
\end{figure}

Pressure was changed and determined by ruby fluorescence at room temperature \cite{Ruby_2,Shen02072020_Ruby} before each run. Low temperature alternating-current (AC) resistance measurements down to 1.8 K were conducted in a Quantum Design physical property measurement system (PPMS), with the Lake Shore AC resistance bridges (models 370 and 372), and a frequency of 17 Hz and 1 mA excitation current.

Room-temperature high-pressure PXRD experiments were conducted at the GSECARS 13-BM-C Beamline of the Advanced Photon Source (APS) at Argonne National Laboratory (ANL). X-rays with a wavelength of 0.4271 Å were focused to a 10 $\mu$m (vertical) × 10 $\mu$m (horizontal) spot size. The polycrystalline sample was ground into powder, and loaded into a wide-opening BX-90 DAC with 300 $\mu$m culet type Ia diamond anvils, a pre-indented Re gasket, and two $<$10 $\mu$m ruby spheres as the pressure manometer. Neon was used as the PTM. Pressure was determined \textit{in situ} using the calibrated pressure-dependence of the $R_1$ line in the ruby fluorescence spectra \cite{Ruby_2,Shen02072020_Ruby}. The two-dimensional diffraction images were integrated using the DIOPTAS software \cite{Prescher03072015_DIOPTAS} and both Rietveld and LeBail refinements were performed in GSAS-\MakeUppercase{\romannumeral 2} \cite{Toby:aj5212_GSAS}.

\section{Data Presentation and Analysis}
\subsection{Ambient Pressure}

\begin{figure*}
\includegraphics[width=17cm]{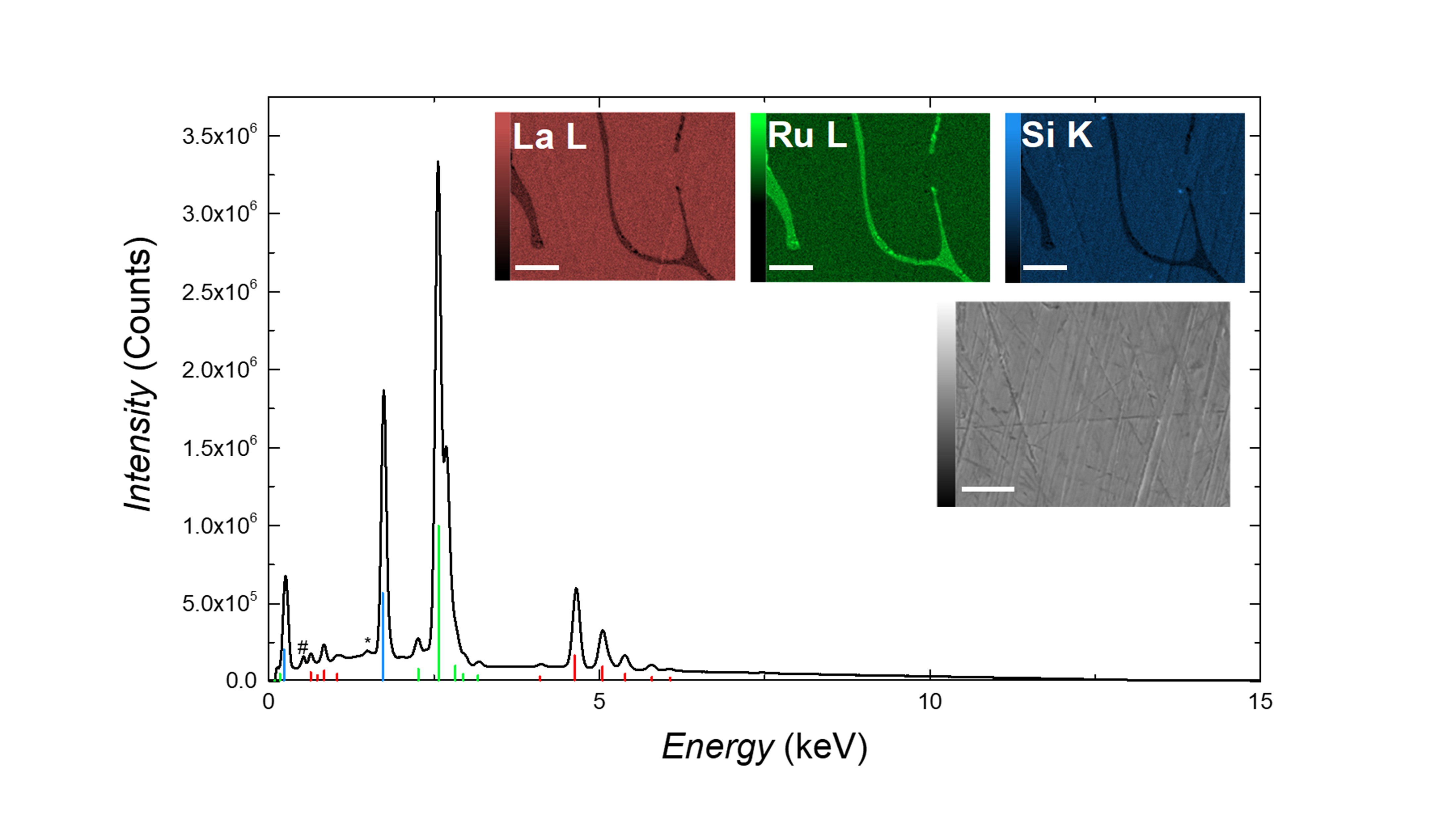}
\caption{\label{fig:eds} Energy dispersive spectroscopy mapping for polished LaRu$_3$Si$_2$. The gray color scale inset show the morphology of the scanned area, while other insets in different color scale shows elemental distribution for each element with the criteria of the characteristic x-ray lines (red for La lines, green for Ru lines, blue for Si lines), with \# and * labeling minor peak from oxygen and aluminum sample holder. Brighter color in each graph denotes higher intensity of x-ray collected. White scale bar stands for 10 $\mu$m.}
\end{figure*}

Figure \ref{fig:PXRD} shows the PXRD pattern of an as-prepared sample at ambient pressure-temperature ($P-T$) conditions, alongside the calculated pattern. Most of the diffraction peaks can be indexed to the hexagonal structure of LaRu$_3$Si$_2$, P6/$mmm$, with lattice parameters $a=5.676$ \r A, $c = 3.558$ \r A, which is consistent with a previous report \cite{lib12a}. Weak impurity peaks were detected from excess Ru and LaRu$_2$Si$_2$. Energy-dispersive spectroscopy (EDS) along with mapping is shown in Fig. \ref{fig:eds}. From the insets of Fig. \ref{fig:eds}, there are several stripe like regions where there is more concentrated Ru, while La and Si show deficiency. Leaving aside such Ru-rich regions, the constituent elements (La, Ru, and Si) are well distributed. Some uniform streaks of La and Si are due to the scratches from polishing, which can be seen in the gray color inset of the sample's morphology. Such inhomogeneity indicates that pure Ru as an impurity phase is not evenly distributed in our sample. We note that the uniform distribution region shows the La:Ru:Si ratio as 18.4$\pm 0.05$: 50.7$\pm0.06$: 30.9$\pm 0.07$, in fair agreement with the main phase LaRu$_3$Si$_2$ stoichiometry (La:Ru:Si = 16.7: 50: 33.3). Considering that we have used \textquoteleft factory standards\textquoteright rather than our own specific ones and given our rather high $RRR$ values, 6-9 (see below), the sharp superconducting transition and the value of $T_c$ consistent with the literature, we thus take that the sample is stoichiometric LaRu$_3$Si$_2$ and consider the studied properties of the sample to be intrinsic. The absence of the LaRu$_2$Si$_2$ phase in EDS and further in high-pressure PXRD (shown in Fig. \ref{fig:HP_PXRD} below) is in contrast with our ambient $P-T$ PXRD result, where a significantly larger amount of the crushed arc-melted sample was examined. Altogether this suggests that LaRu$_2$Si$_2$ could be more irregularly dispersed.

\begin{figure}[b]
\includegraphics[width=8cm]{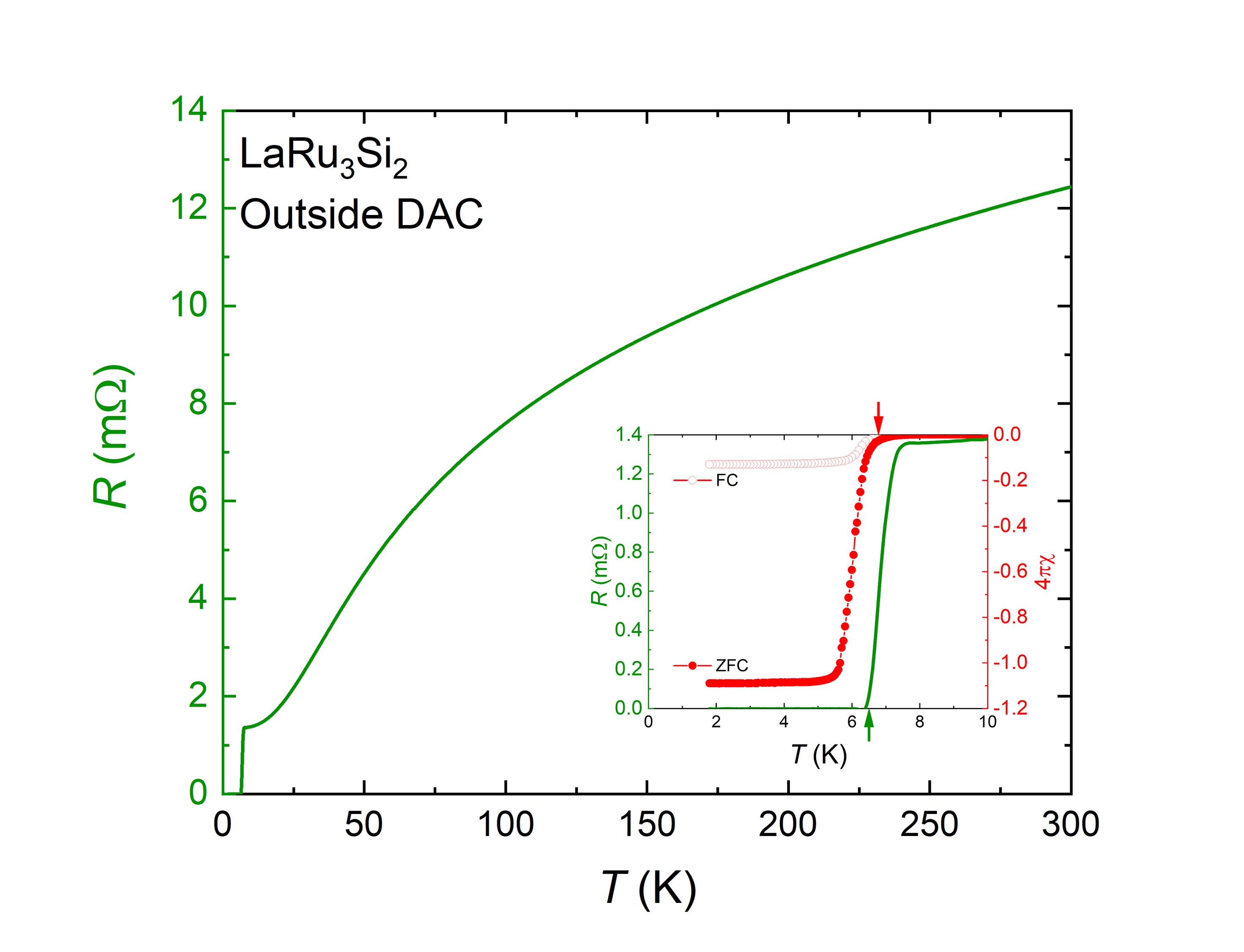}
\caption{\label{fig:ambient} Temperature-dependent electrical resistance of sample outside the DAC from 1.8 to 300 K; the inset shows the temperature-dependent ZFC, FC magnetization measured in a 20 Oe magnetic field, and resistance from 1.9 to 10 K. Criteria of superconducting transition as offset of resistance and onset of magnetization (ZFC) measurement are indicated by green and red arrows, respectively.}
\end{figure}

\begin{figure*}[t]
\includegraphics[width=5.5cm]{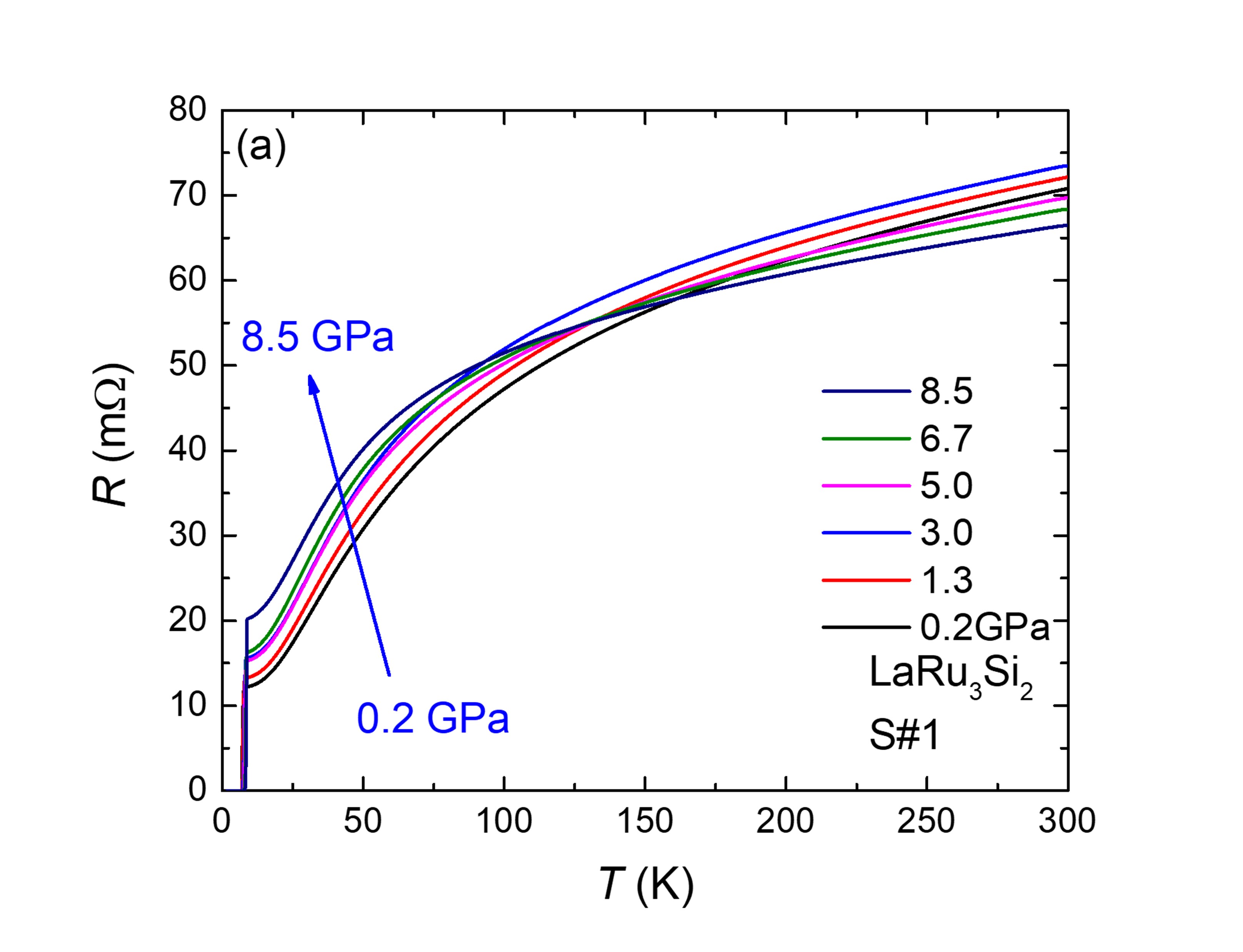}
\includegraphics[width=5.5cm]{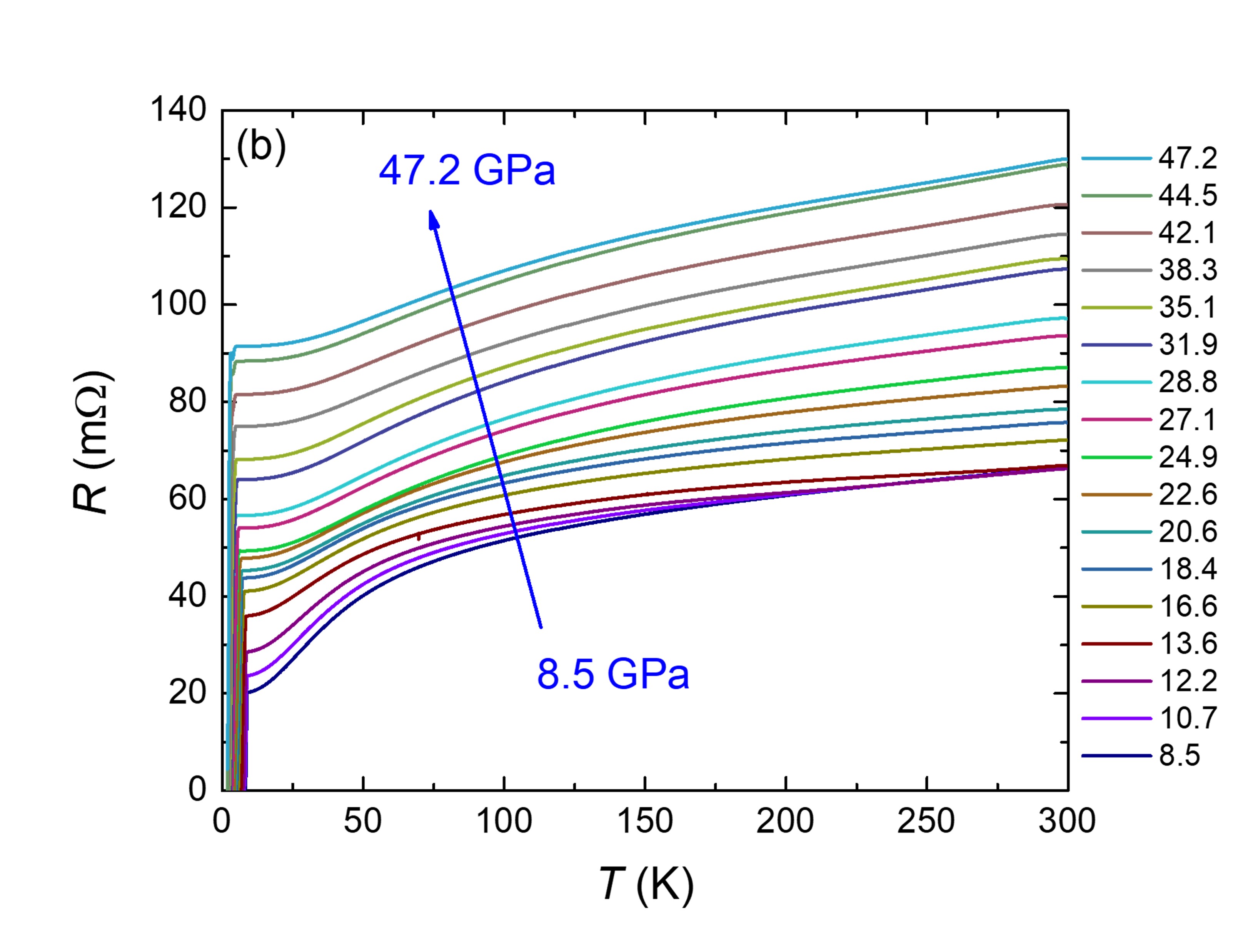}
\includegraphics[width=5.5cm]{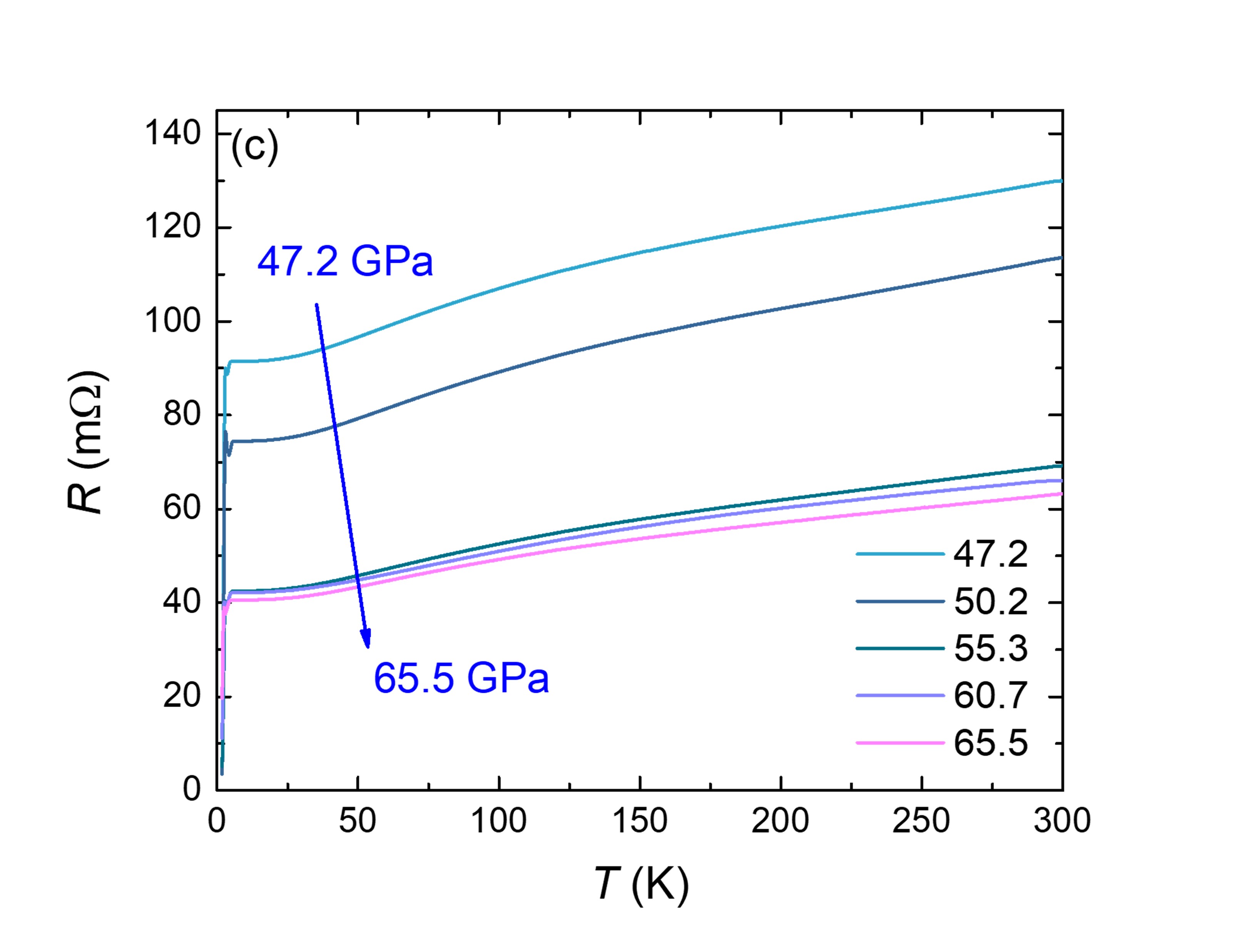}
\includegraphics[width=5.5cm]{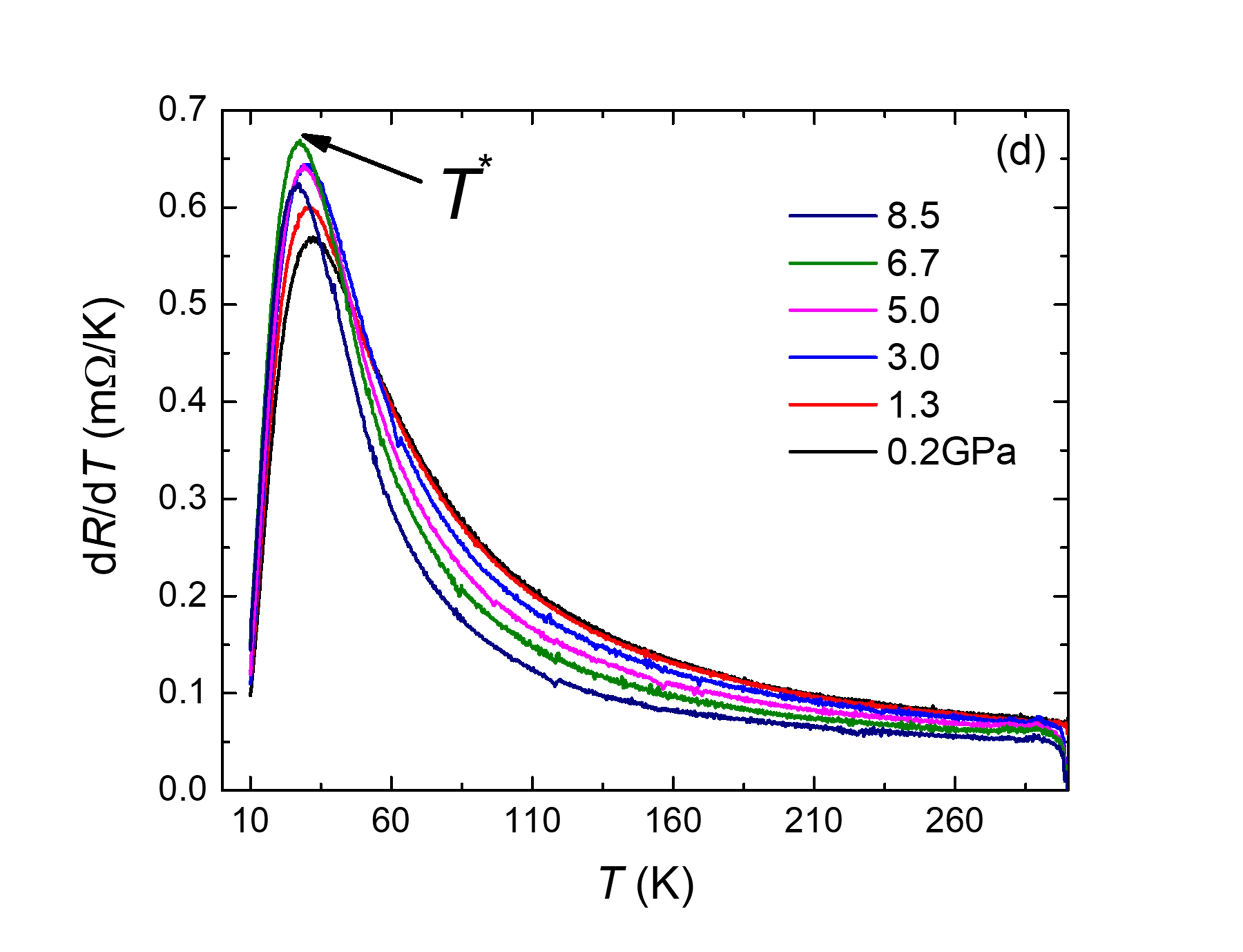}
\includegraphics[width=5.5cm]{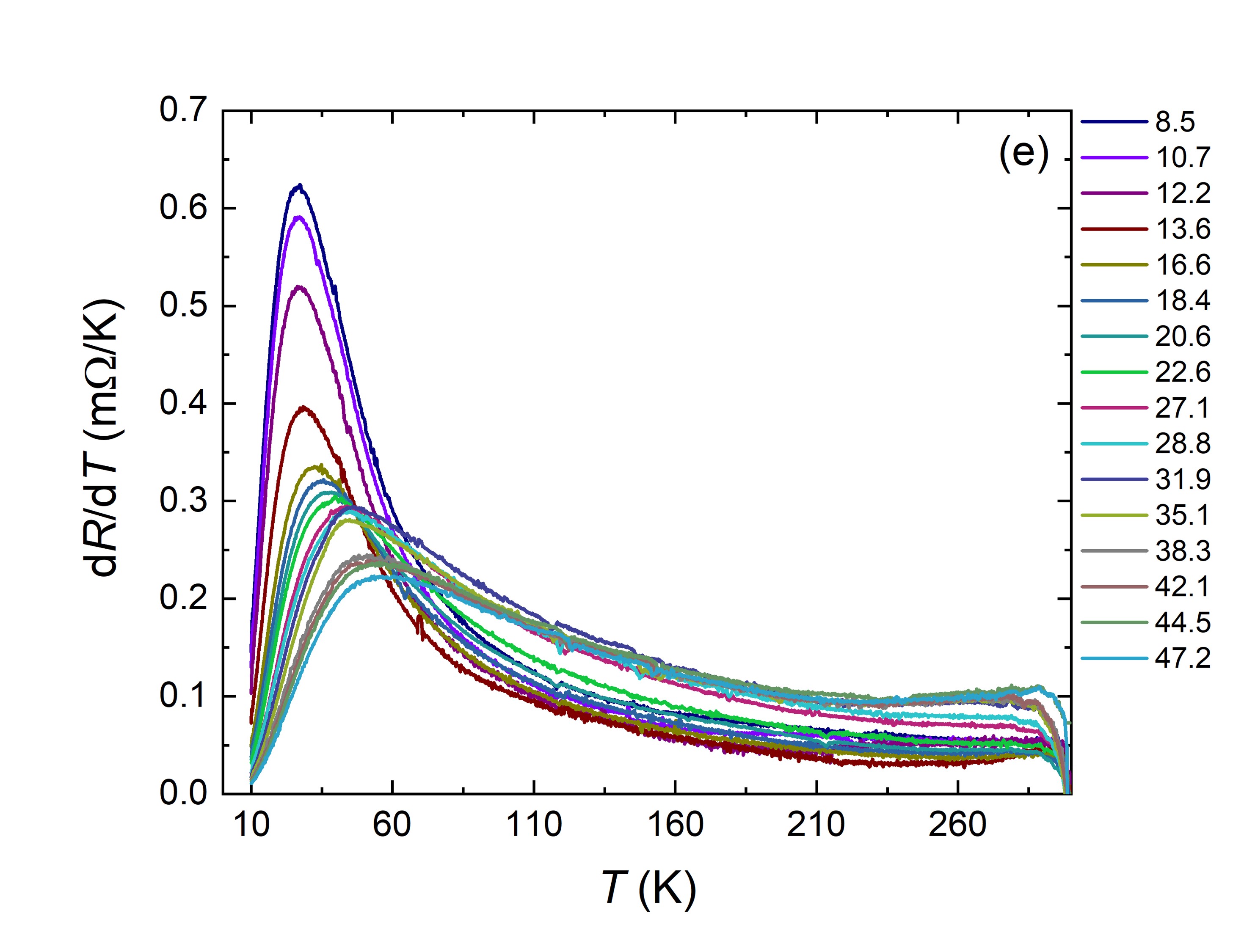}
\includegraphics[width=5.5cm]{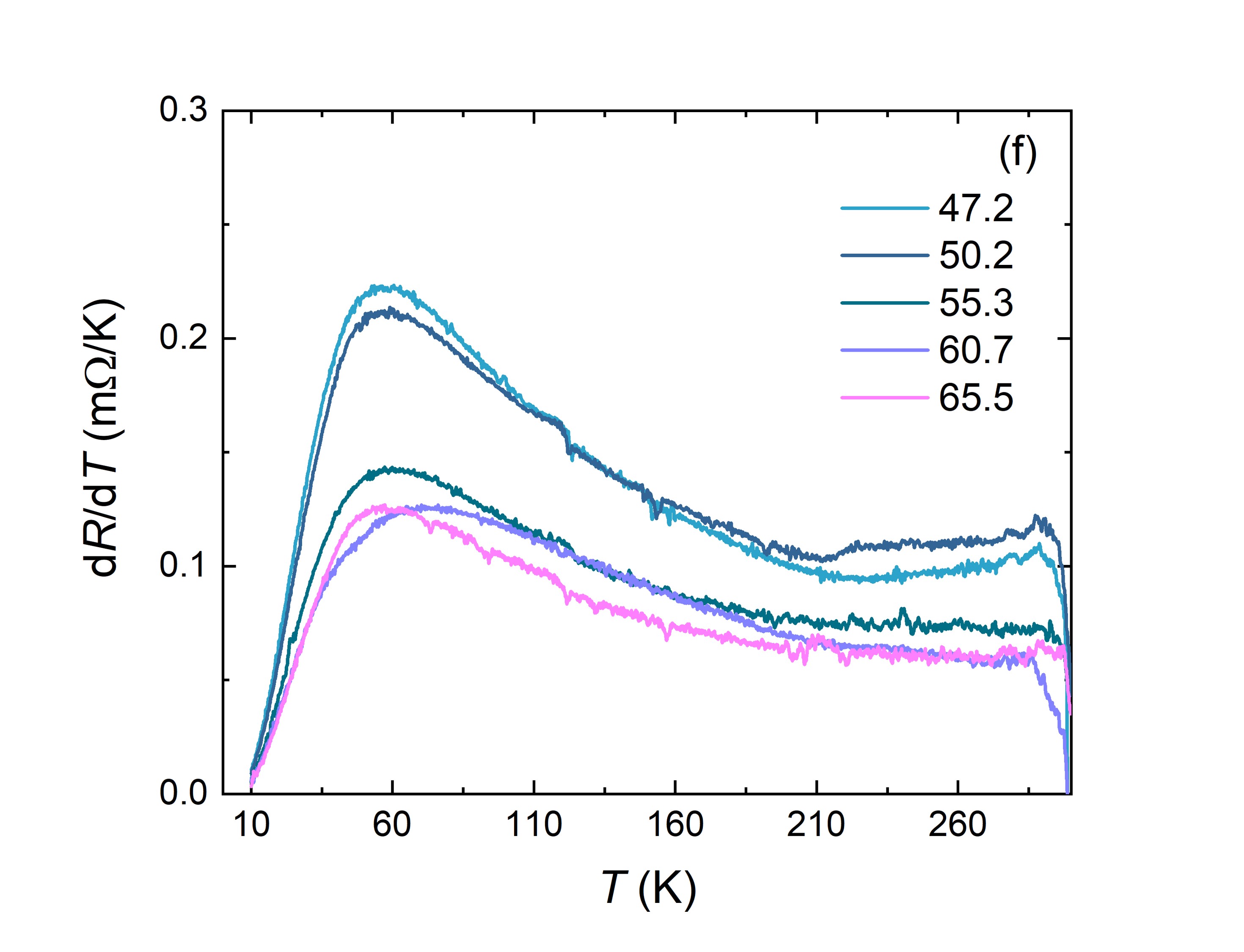}
\caption{\label{fig:RT_300K}(a)--(c) Temperature-dependent  electrical resistance dependence of S\#1 at various pressures over 1.8 - 300 K; (d)--(f) Temperature derivatives of the resistance ($<$10 K range skipped for clarity). Pressure range of (a), (d) 0.2 - 8.5 GPa, (b), (e) 8.5 - 47.2 GPa, (c), (f) 47.2 - 65.5 GPa.}
\end{figure*}

\begin{figure*}[t]
\includegraphics[width=5.5cm]{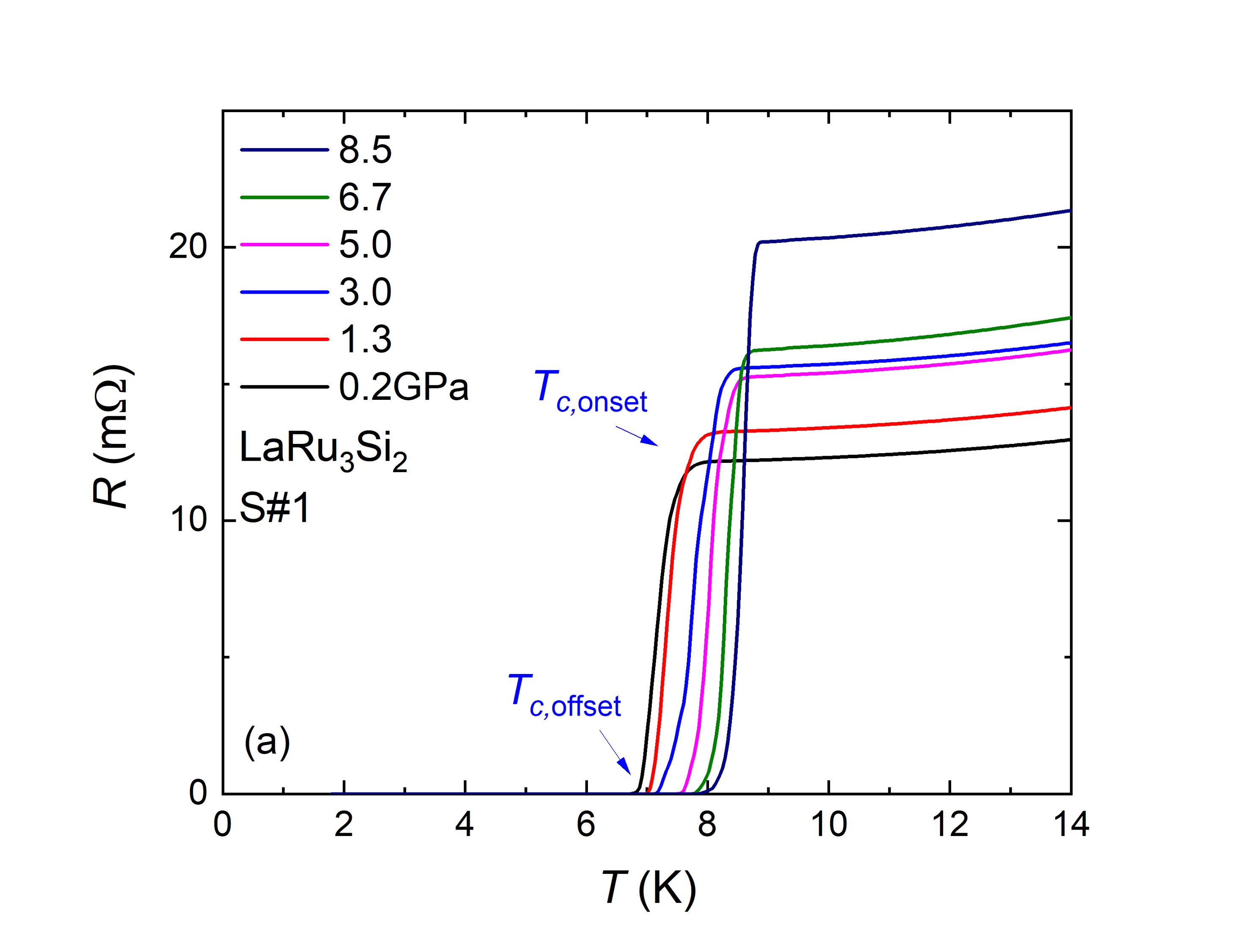}
\includegraphics[width=5.5cm]{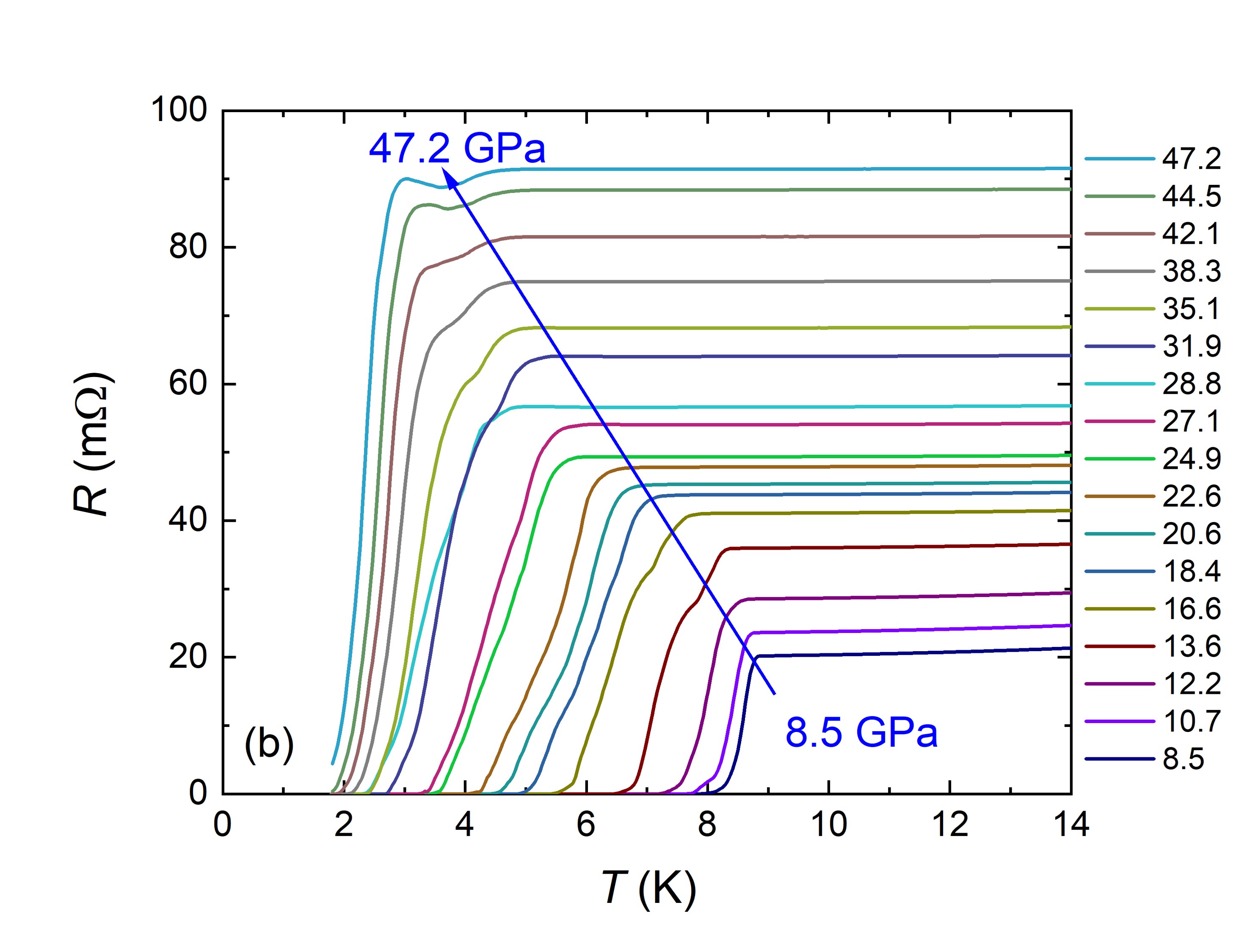}
\includegraphics[width=5.5cm]{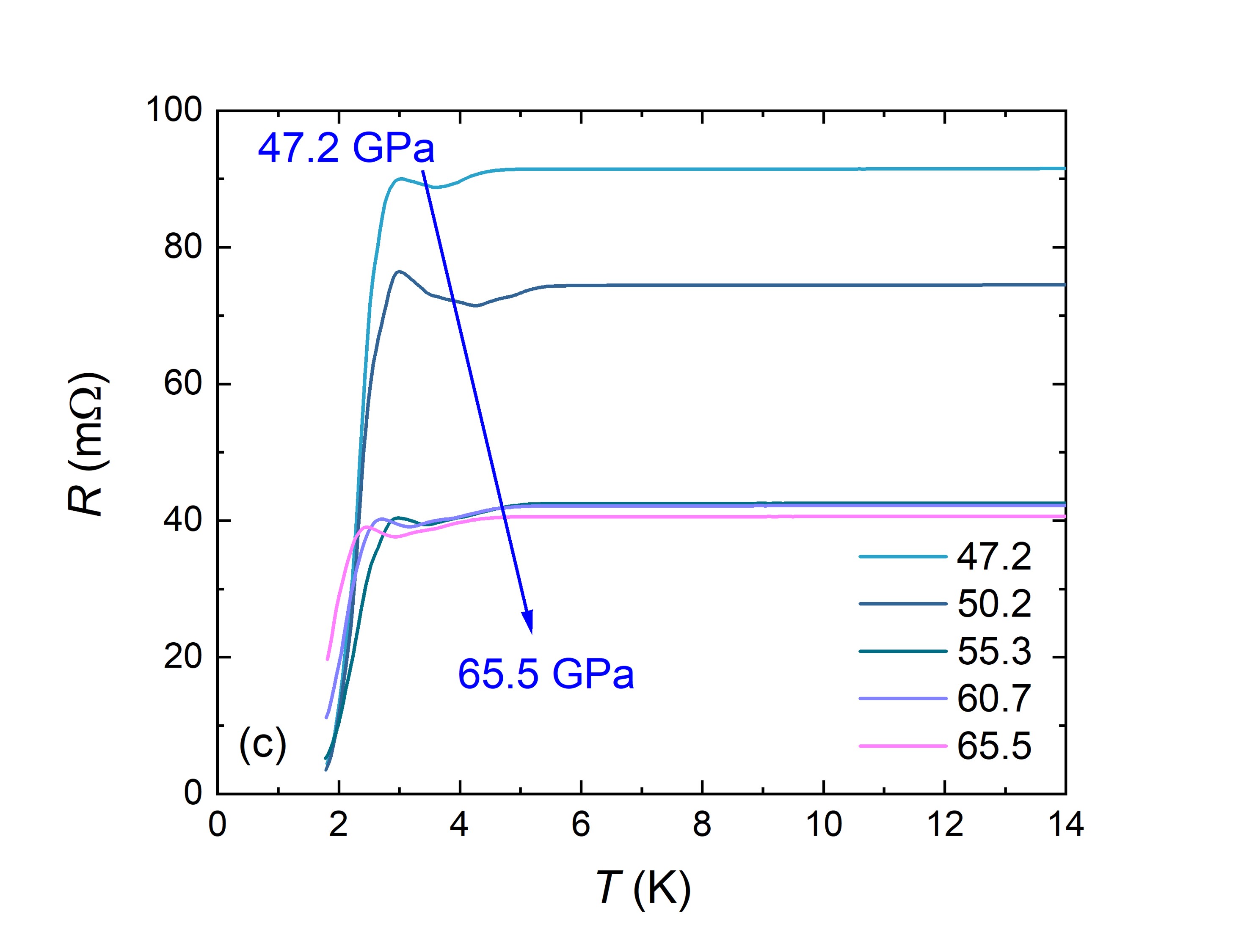}
\caption{\label{fig:RT_varP} (a)--(c) Low-temperature (1.8 - 14 K) part of the electrical resistance at different pressures. }
\end{figure*}

To characterize the superconducting (SC) state, temperature-dependent magnetization measurement and electrical resistance measurement were performed at ambient pressure. Figure \ref{fig:ambient} shows the zero-field resistance measurement from 1.8 K to 300 K, revealing metallic behavior of decreasing resistance upon cooling along with a sharp superconducting transition starting at 7.1 K along with an unusual shoulder like curve shape in good agreement with the previous report \cite{mie21a}. Figure \ref{fig:ambient} inset presents the ZFC and FC magnetization measurements of a three-dimensional, irregularly shaped sample in an applied field of 20 Oe. The susceptibility $\chi$ was calculated by approximating a spherical sample geometry; thus the demagnetization factor was taken as $1/3$, $\chi =  \cfrac{M}{H-4\pi M/3}$. At 1.8 K, ZFC susceptibility reaches -1.09, close to -1, consistent with the $\sim$100\% volume fraction of the as-prepared LaRu$_3$Si$_2$ sample, although there remains a small amount of secondary phases of Ru and LaRu$_2$Si$_2$. The offset of the superconducting transition in $R(T)$ is determined by the intersection of maximum slope and $T=0$ axis, whereas the onset of superconducting transition in $4\pi\chi(T)$ is determined based on the initial deviation of the $4\pi\chi$ curve from $\chi = 0$ starts, as indicated in the Fig. \ref{fig:ambient} inset. Both of them agree well and give $T_c = 6.6 \pm 0.2$ K, which is consistent with $T_{c,\text{offset}}$ in a previous report \cite{anomalous_11}.

\subsection{Resistance under Pressure}
We performed high-pressure resistance measurements on two samples from the same batch, denoted as samples S\#1 and S\#2. Sample S\#1 was measured with higher pressure density up to 65.5 GPa, whereas S\#2 was measured up to 60.5 GPa. Both samples show a SC dome under pressure but with minor differences at high pressure. A detailed discussion of the resistance measurements of the two samples can be found in Appendix A. It should be noted that upon cooling, the DAC could experience pressure change due thermal contraction of cell components. Based on the previous report of a similar design DAC \cite{Gav09_dPdT}, we expect not more than 1-2 GPa increase of pressure in the cell upon cooling.
\begin{figure*}
\includegraphics[width=5.5cm]{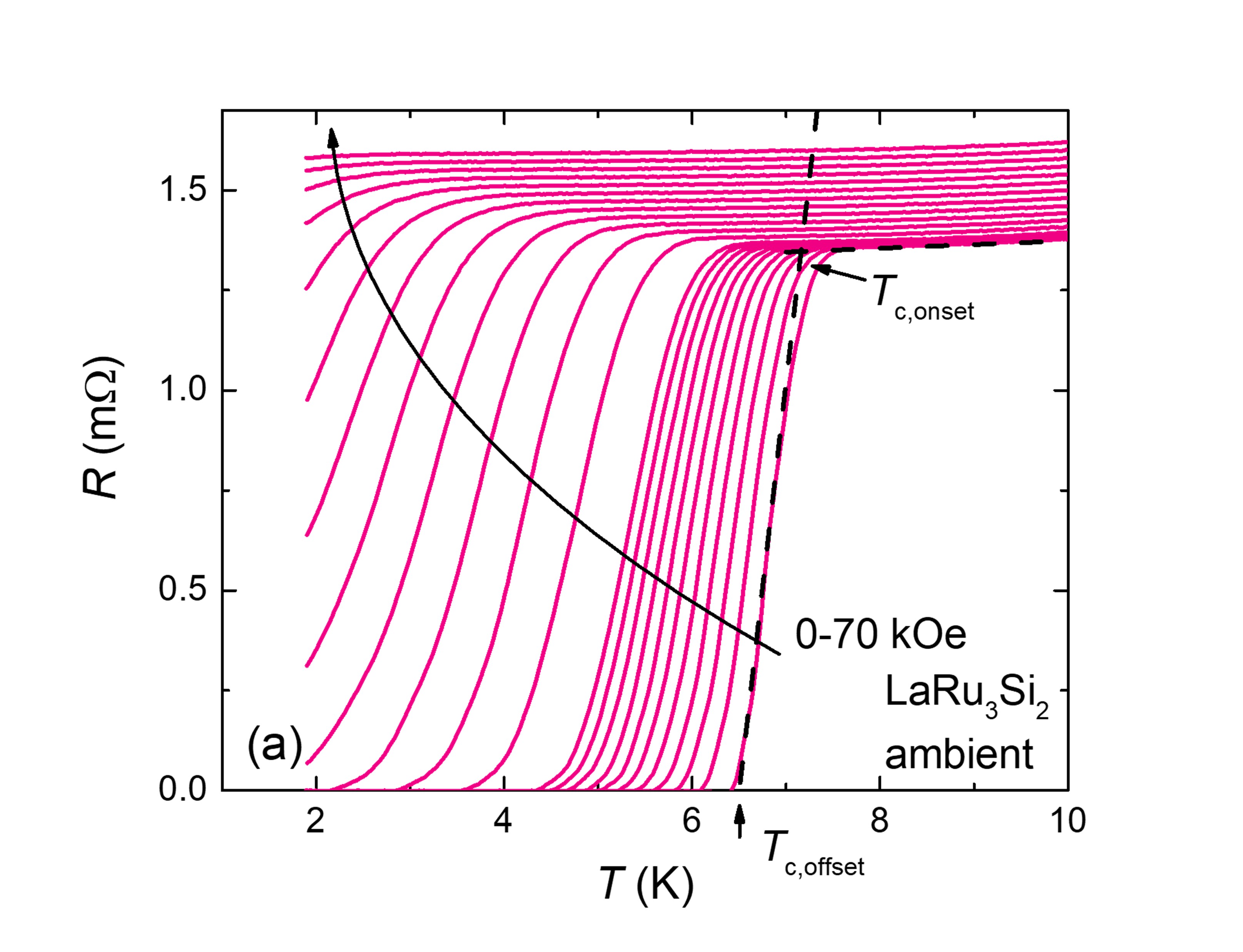}
\includegraphics[width=5.5cm]{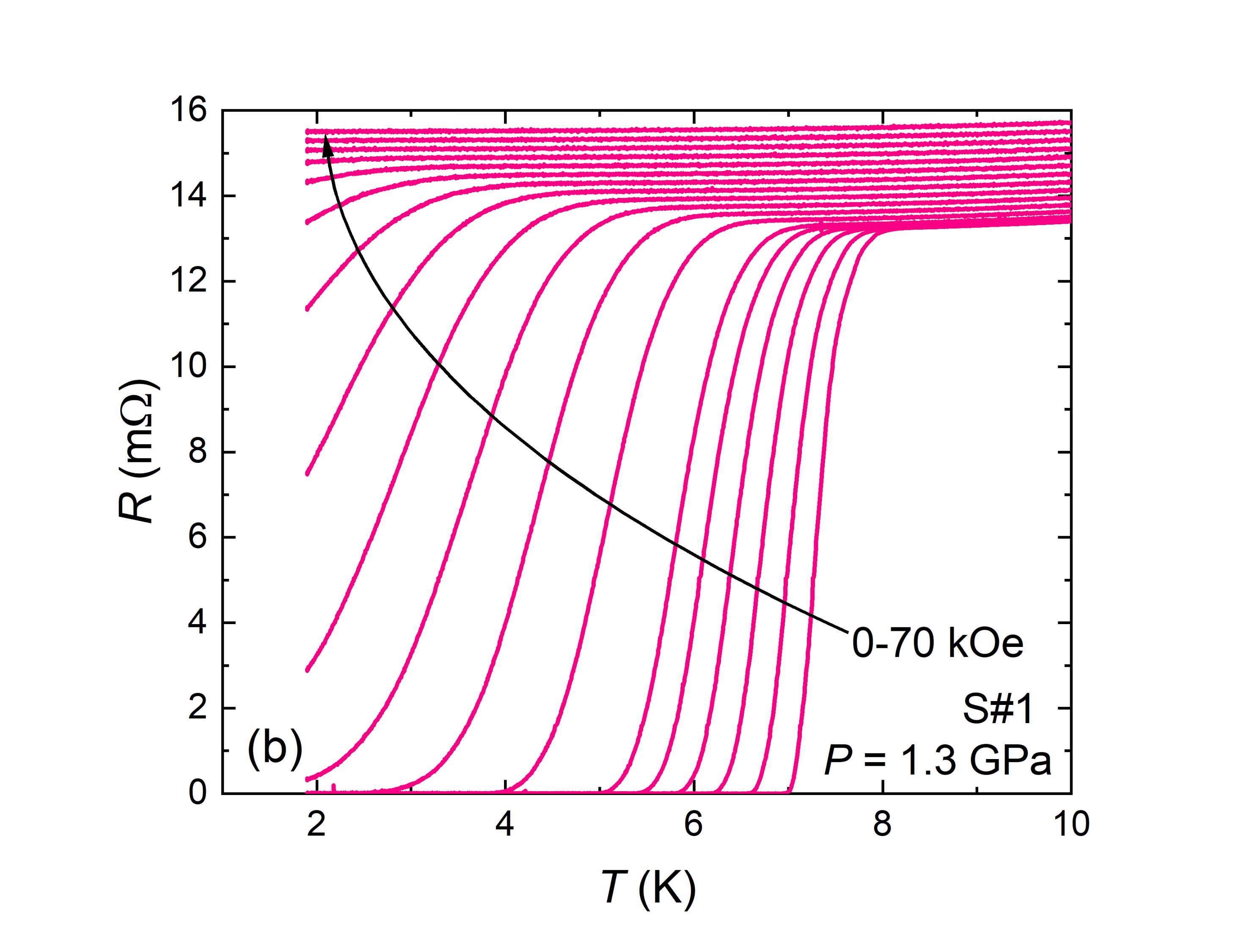}
\includegraphics[width=5.5cm]{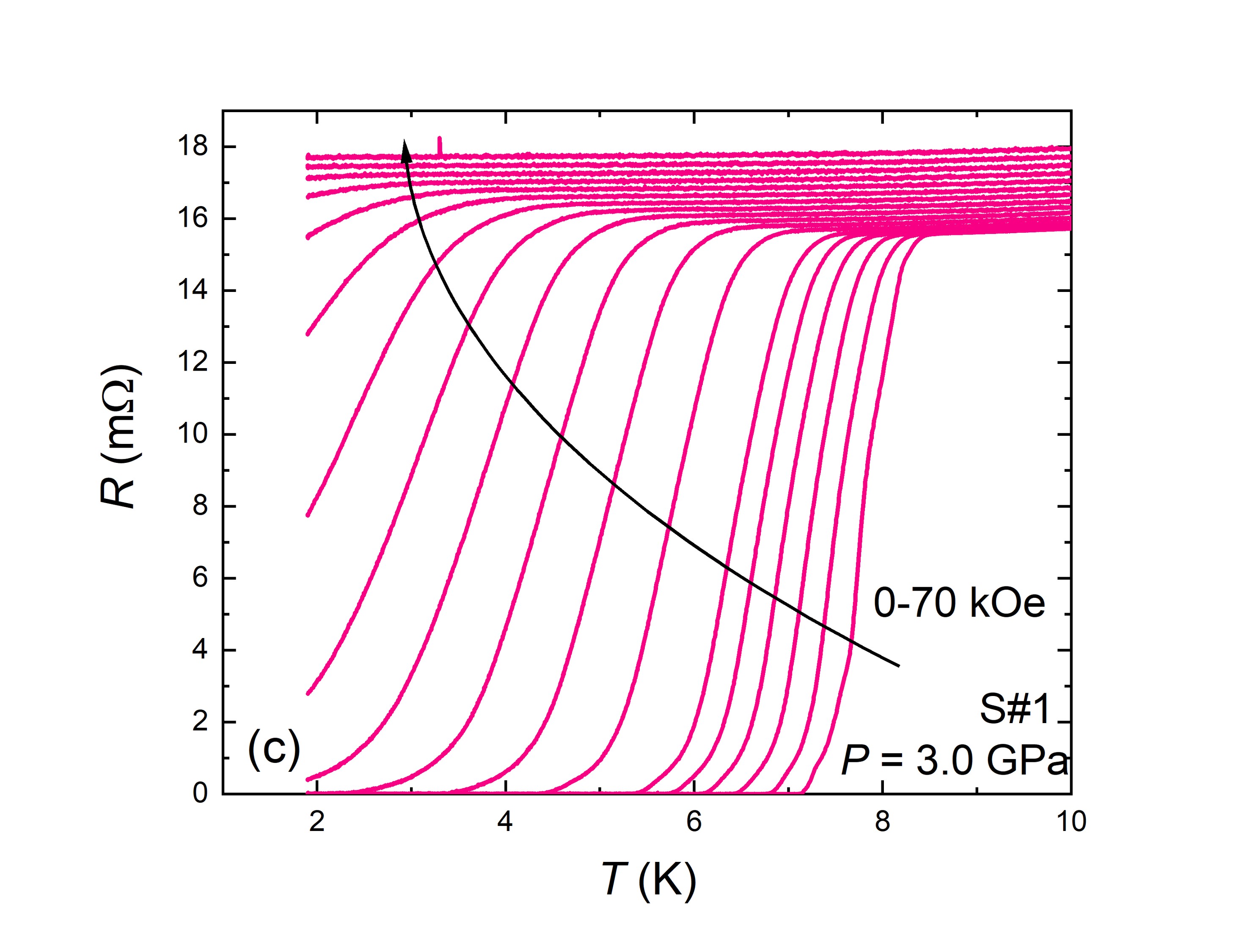}
\includegraphics[width=5.5cm]{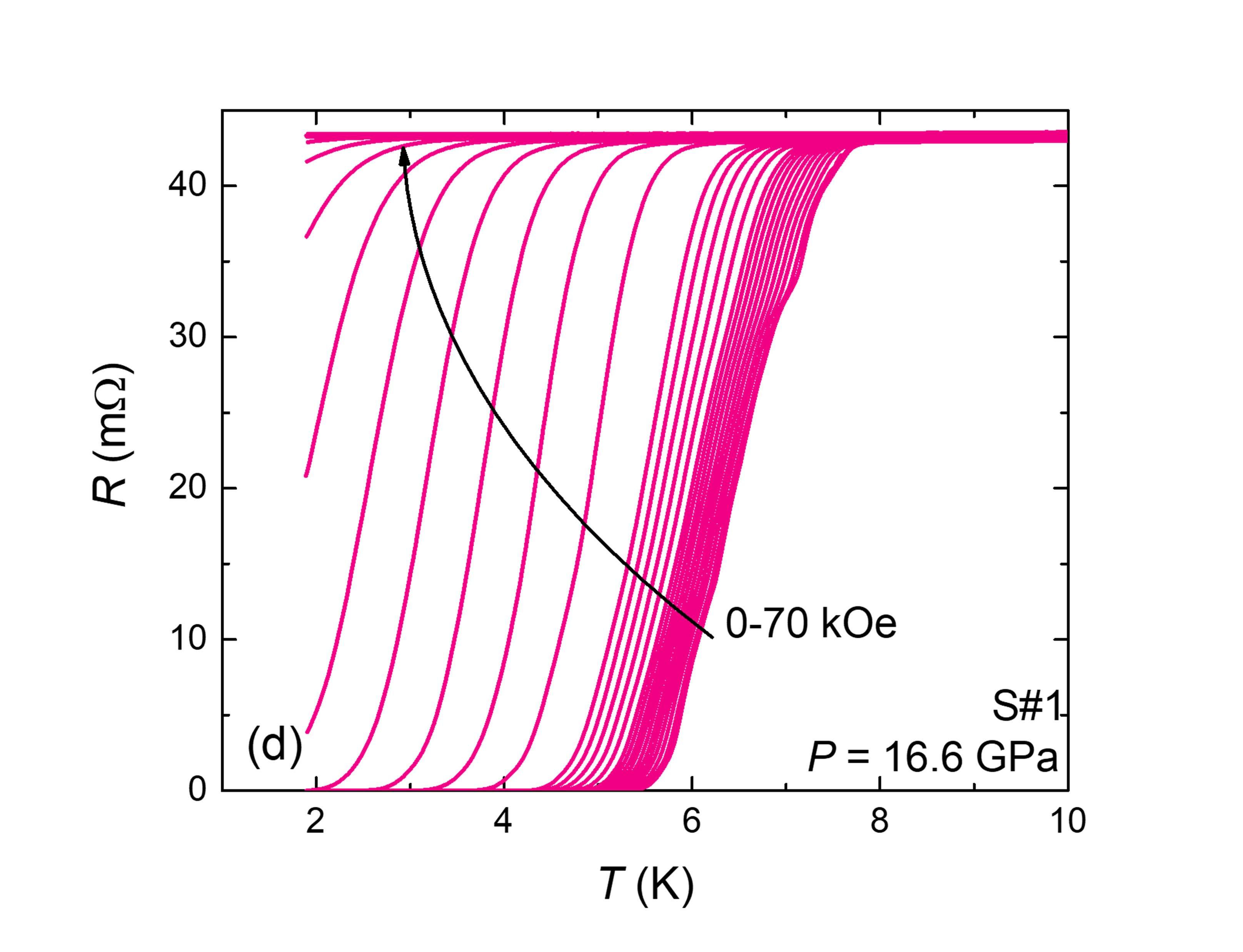}
\includegraphics[width=5.5cm]{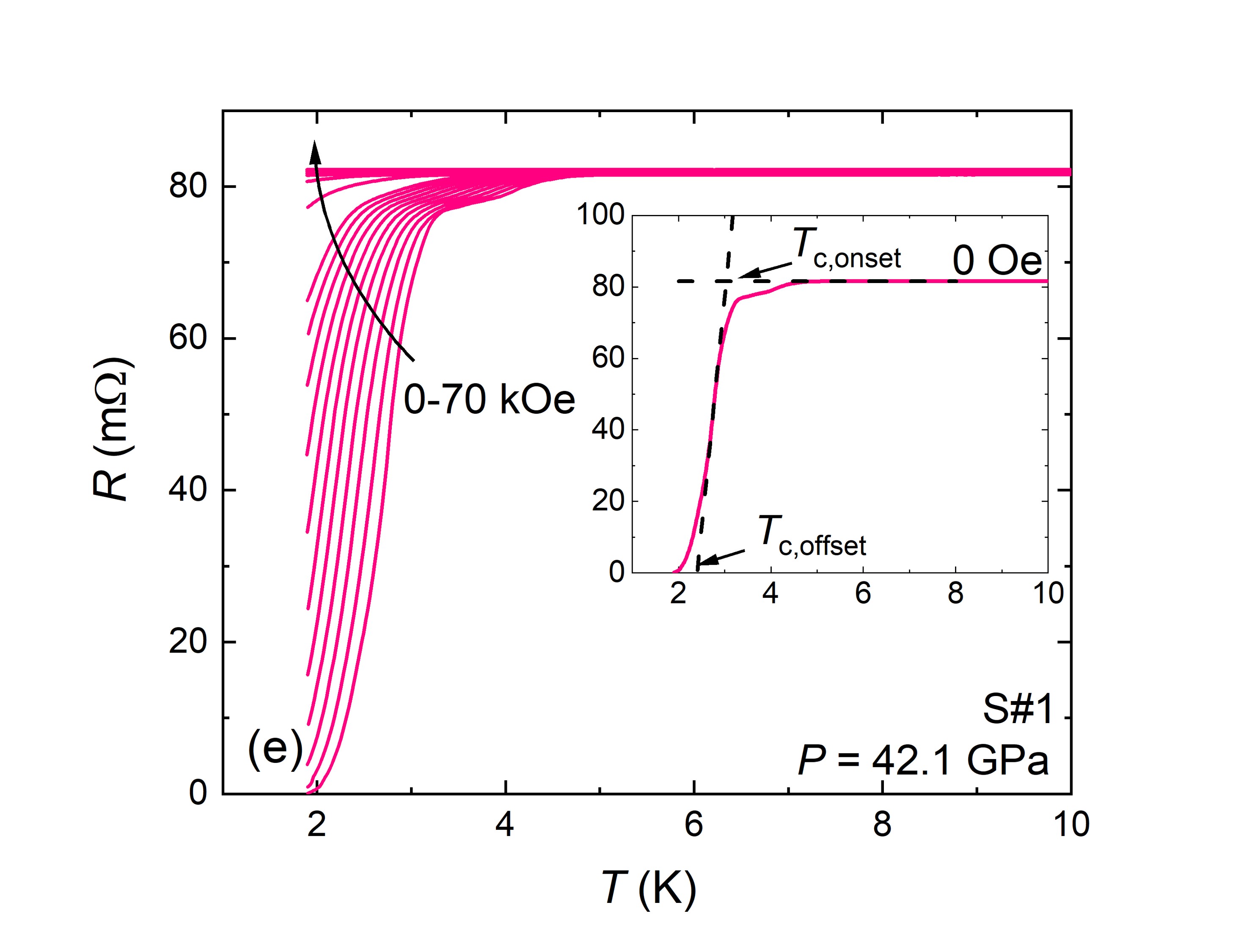}
\includegraphics[width=5.5cm]{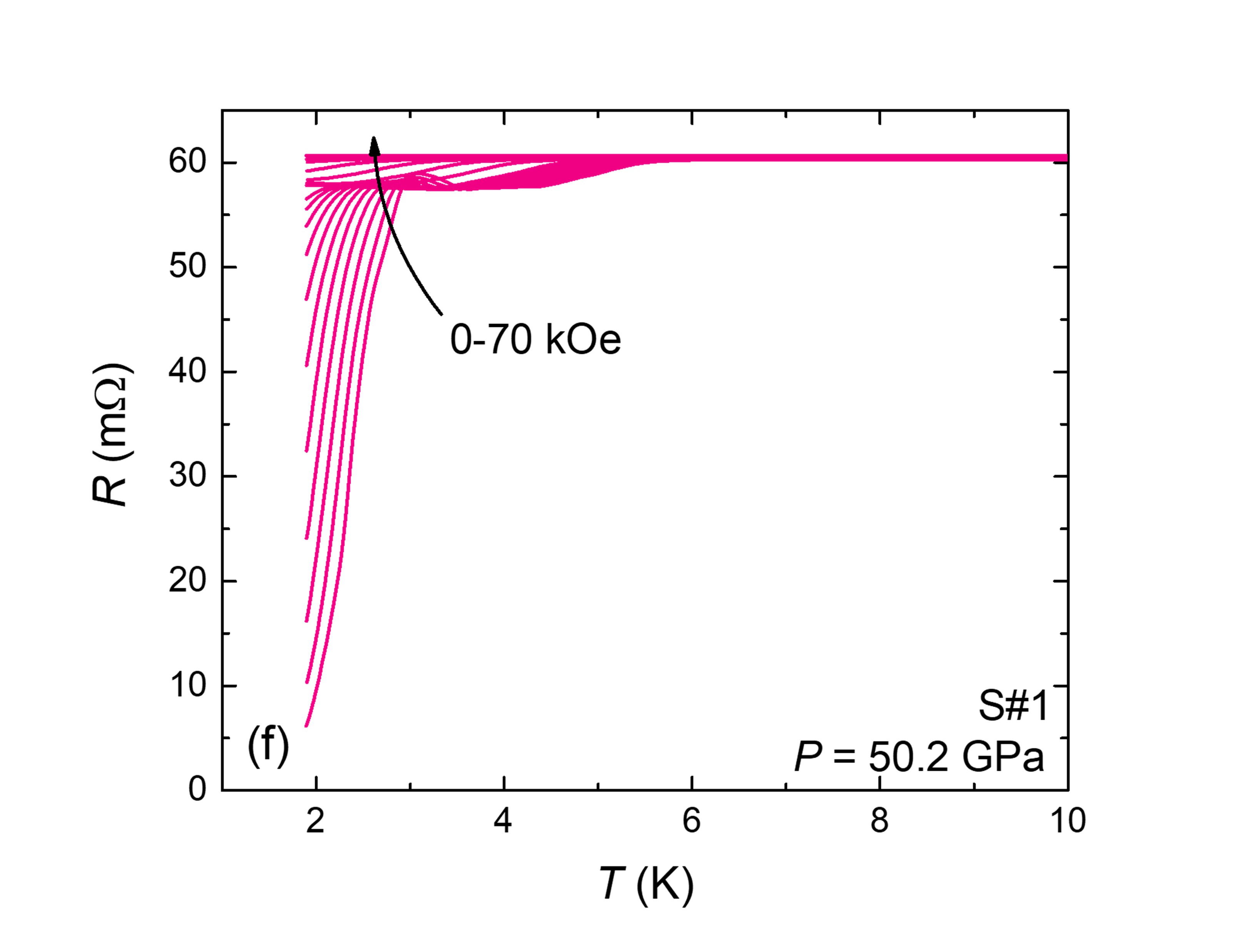}
\caption{\label{fig:RT_varH_varP} (a-f) Electrical resistance dependence on temperature from 1.9 to 10 K, at various pressure under magnetic field from 0 to 70 kOe. Criteria of $T_c$ onset and offset at 0 Oe is shown in (a) and (e).}
\end{figure*}
Temperature-dependent resistance, $R(T)$, data from 0.2 GPa to 65.5 GPa of sample S\#1 are shown in FIG. \ref{fig:RT_300K} (a-c), ranging from 1.8 K to 300 K. For clarity, we present our results over three panels. 
Whereas for pressures up to 8.5 GPa there is relatively little change in the $R(T)$ curves (FIG. \ref{fig:RT_300K} (a)), for pressures above 8.5 GPa there is a clear change in the size of the resistance, the shape of the $R(T)$ plot and the value of the residual resistivity ratio ($RRR = R(300K)/R_0$). The residual resistance $R_0$ is taken as the intercept from a linear fit of the $R(T)$ plateau before the superconducting transition. With further increasing pressure from 47.2 GPa to 65.5 GPa, shown in FIG. \ref{fig:RT_300K} (c), the $R(T)$ curve shifts downward with an abrupt drop from 55.3 GPa to 60.7 GPa, and the curvature in the normal state becomes less pronounced. Note that such a drop motivated the measurement of sample S\#2 and was not reproduced. As such, this anomaly is most likely associated with some combination of sample and contact damage and/or shifting rather than being an intrinsic property of LaRu$_3$Si$_2$.

The derivatives of temperature-dependent resistance, ${dR}/{dT}$, from 10 K to 300 K are presented in FIG. \ref{fig:RT_300K} (d-f) to characterize the shoulder like behavior of resistance. A peak is observed over the whole pressure range. $T^*$ is then defined as the peak position of ${dR}/{dT}$ before the superconducting transition, i.e., $>$ 10 K. Figure \ref{fig:RT_300K} (d) shows an arrow for $T^*$ at 6.7 GPa. This peak tends to broaden upon increasing pressure, with $T^*$ shifting to lower temperature first and then increasing. Previously, this anomaly temperature was identified as being where the sign changes for Hall signal from positive (holes) to negative (electrons), and was argued to be indicating a complex interplay of electronic states \cite{mielke2024chargeordersdistinctmagnetic}. We found the evolution of $T^*$ is correlated with $T_c$, which will be further discussed after presenting the $T-P$ phase diagram in Fig. \ref{fig:TC_varP} below.

Figure \ref{fig:RT_varP} (a-c) present the low-temperature ($T \leq 14$ K) resistance versus temperature data over the same pressure ranges as in Fig. \ref{fig:RT_300K}. In Fig. \ref{fig:RT_varP} (a), we define the criteria of superconducting transition temperature onset $T_{c,\text{onset}}$ as the intersection between the maximum slope and the upper plateau before the superconducting transition, and the offset of the superconducting transition $T_{c,\text{offset}}$ as the intersection between the maximum slope and the $T=0$ axis. At pressures where the superconducting transition exhibits multi step-like behavior (see below), only the sharpest transition is taken as the characteristic transition for maximum slope.

In Fig. \ref{fig:RT_varP} (a), from 0.2 GPa to 8.5 GPa, both $T_{c,\text{onset}}$ and $T_{c,\text{offset}}$ increase by more than 1 K.  $T_{c,\text{onset}}$ and $T_{c,\text{offset}}$ are then suppressed with more pressure being applied, as shown in Fig. \ref{fig:RT_varP} (b,c). It is also worth noting that the original, single, sharp SC transition gets broader and evolves to be multi step-like, resulting in a hump like curve beyond 44.5 GPa. This has been reported of the good connection between SC grains of arc-melted LaRu$_3$Si$_2$ \cite{anomalous_11, RhIr_Doping_2023}, despite the presence of minimal secondary phases. Such multi step-like behavior could be due to be some degree of stoichiometric variation and/or pressure inhomogeneity.

To further study the superconducting state under pressure, we performed electrical transport measurement under external magnetic field varying from 0 to 70 kOe at several pressures on sample S\#1, along with ambient pressure measurements of another sample from the same batch outside the DAC (Fig. \ref{fig:RT_varH_varP} (a-f)). For the pressures in Fig. \ref{fig:RT_varH_varP} (a-c), the superconducting transition remains sharp, with a single drop to zero resistance. In Fig.\ref{fig:RT_varH_varP} (d-f), the superconducting transition shows a multi step feature at zero field.

The upper critical field, $H_{c2}$, is extracted from Fig. \ref{fig:RT_varH_varP} (a-f). As shown in Fig. \ref{fig:RT_varH_varP} (a,e), the onset and offset of the superconducting transition is found by the same aforementioned criteria at different pressures, with $T_{c,\text{onset}}$ derived from sharpest transition, since other small steps get easily suppressed under field.

We can fit $H_{c2}$ versus $T_{c,\text{onset}}$ (presented in Fig. \ref{fig:Hc2_varP}) and $T_{c,\text{offset}}$ (shown in Appendix B) by both a linear fit and the Werthamer–Helfand–Hohenberg (WHH) model \cite{WHH_1966_1, WHH_1966_2, WHH_1966_3}. Details of the fitting method and results can be found in Appendix B, where at each measured pressure the upper critical fields are well below the Pauli paramagnetic limit $H_P=18.3T_c$ (kOe) \cite{Tinkham_SC}. It is evident that, at least for $P$ = 0, 1.3, 3.0, and 16.6 GPa, the WHH model does not accurately describe the experimental data. In addition, from the inset of Fig. \ref{fig:Hc2_varP}, -$\cfrac{dH_{c2}}{dT}(T = T_c)$ is nearly pressure-independent, ranging from 9.2 kOe/K at ambient to 11.4 kOe/K at highest pressure. The -$\cfrac{dH_{c2}}{dT}$ values extracted from the offset data are dramatically different (as shown in Appendix B, Fig.\ref{fig:Hc2_varP_offset} inset), showing dramatic pressure sensitivity with a local maximum near 10 GPa. The origin of this difference is not clear at this point in time.
 
\begin{figure}[h]
\includegraphics[width=8cm]{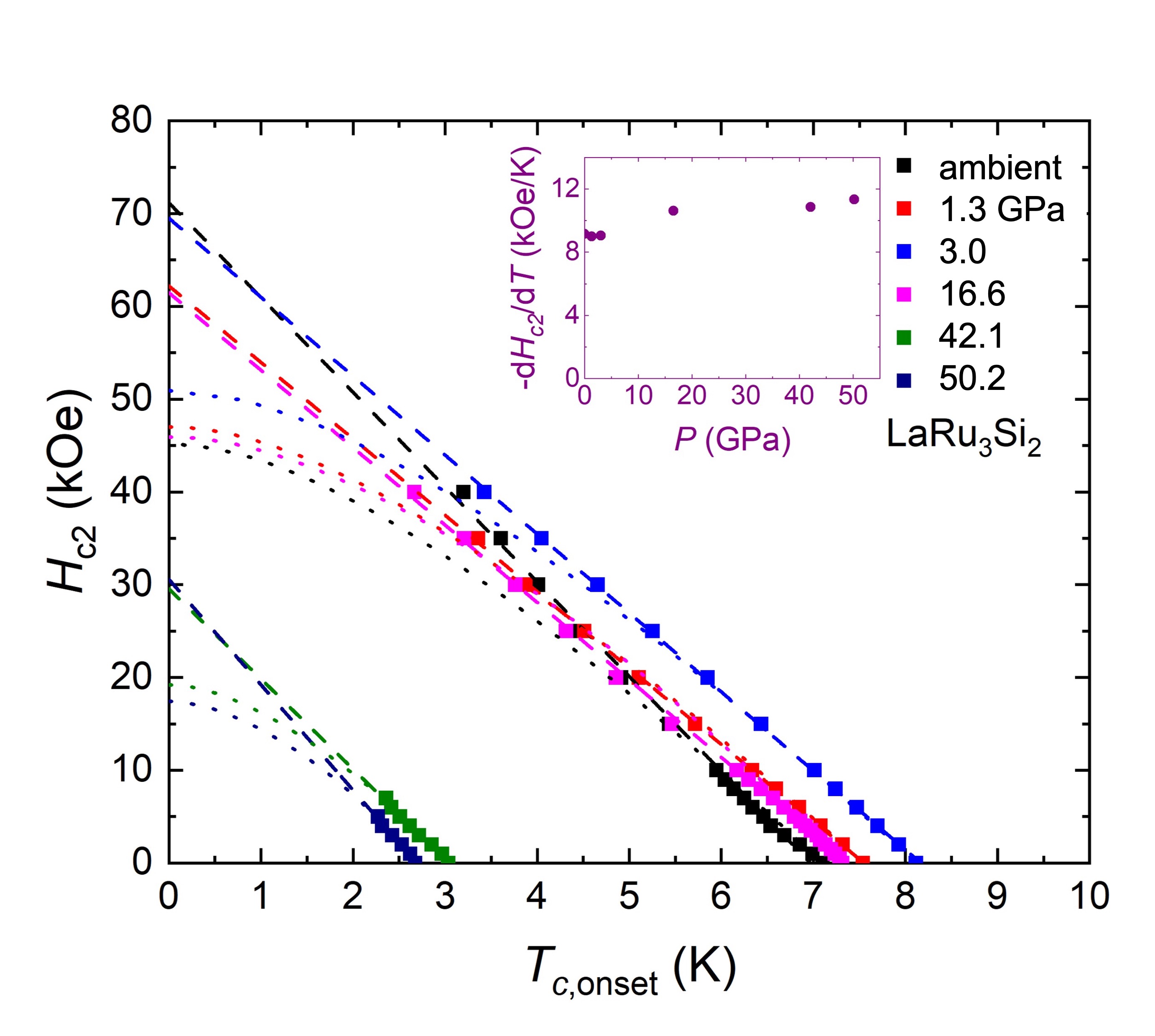}
\caption{\label{fig:Hc2_varP} Upper critical field $H_{c2}$ dependence of $T_{c,\text{onset}}$ obtained from transport measurements at several pressures, inset showing the temperature derivative of $H_{c2}$ at $T_{c,\text{onset}}$. Dashed and dotted lines are the fitted $T$ linear lines and WHH fitted lines for each pressure, respectively. %dotted lines are fitted WHH model lines.
}
\end{figure}

\subsection{PXRD under Pressure}

To investigate the nature of dome like superconductivity in pressurized LaRu$_3$Si$_2$, we performed high-pressure PXRD measurements up to 34.9 GPa. Figure 8 presents the diffraction pattern at the lowest measured pressure (0.6 GPa), with the peaks of the primary phase, LaRu$_3$Si$_2$, indexed. Subtle peaks corresponding to elemental Ru, identified as an impurity phase, are also observed. 
However, the intensities of LaRu$_3$Si$_2$ peaks are not consistent with ambient pressure measurement outside the DAC with significantly larger amount of the sample; i.e., whereas the [200] peak shows maximum intensity for the ambient pressure measurement, the [201] peak shows maximum intensity for PXRD inside the DAC. This brings the consideration of developing preferred crystalline orientation for peak intensity evolution under pressure.
Notably, the impurity phase LaRu$_2$Si$_2$, detected in the ambient-pressure XRD measurements (Fig. \ref{fig:PXRD}), was absent under high-pressure conditions. Missing any noticeable amount of LaRu$_2$Si$_2$ could be due to the inhomogeneity of how the small amount of it is distributed throughout the arc-melted button (consistent with its absence in the EDS, as discussed above).

\begin{figure}[h]
\includegraphics[width=8cm]{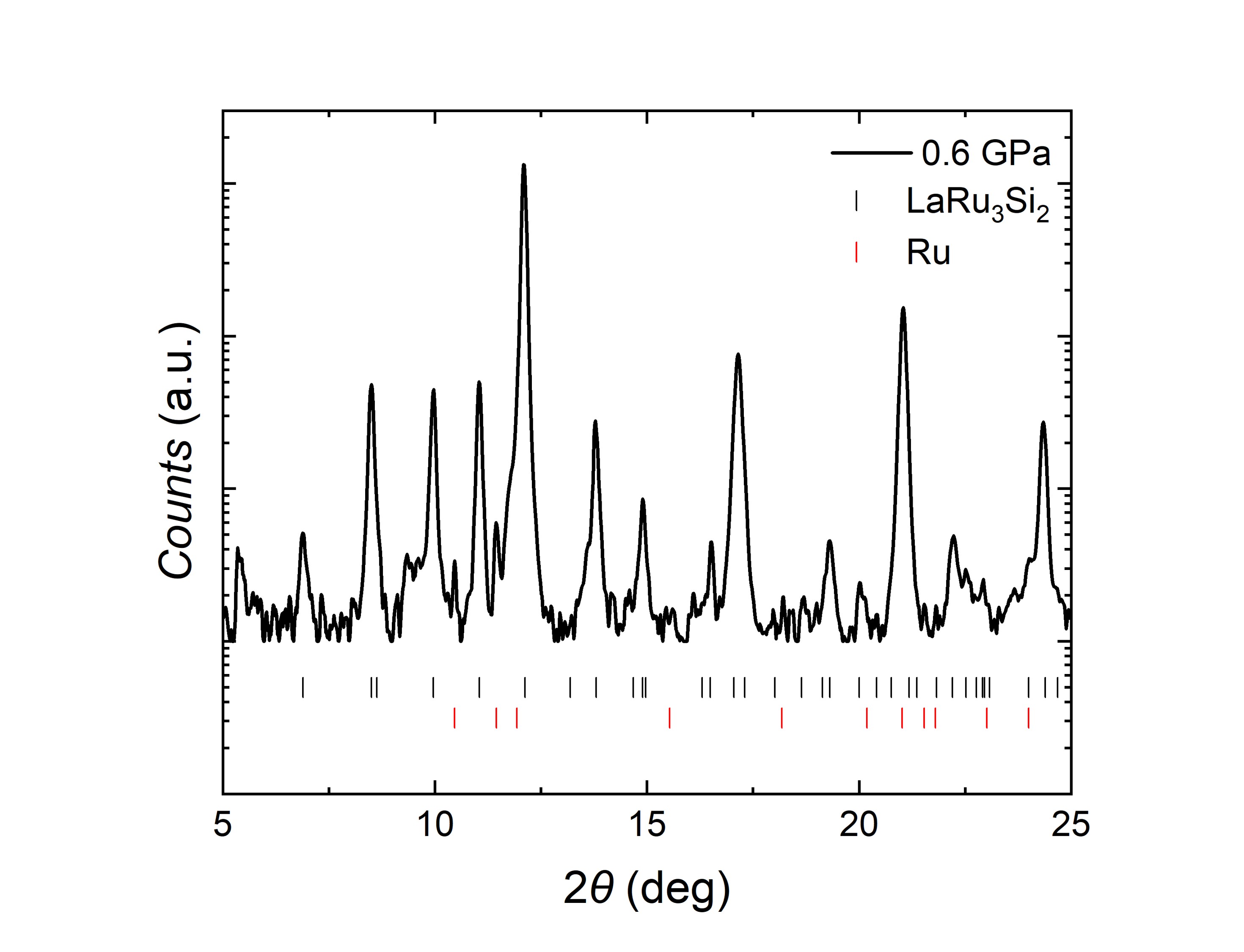}
\caption{\label{fig:0.6GPa_PXRD} Room-temperature PXRD patterns of LaRu$_3$Si$_2$ at 0.6 GPa in semilog scale, with all main peaks well indexed.}
\end{figure}

\begin{figure}[h]
\includegraphics[width=8cm]{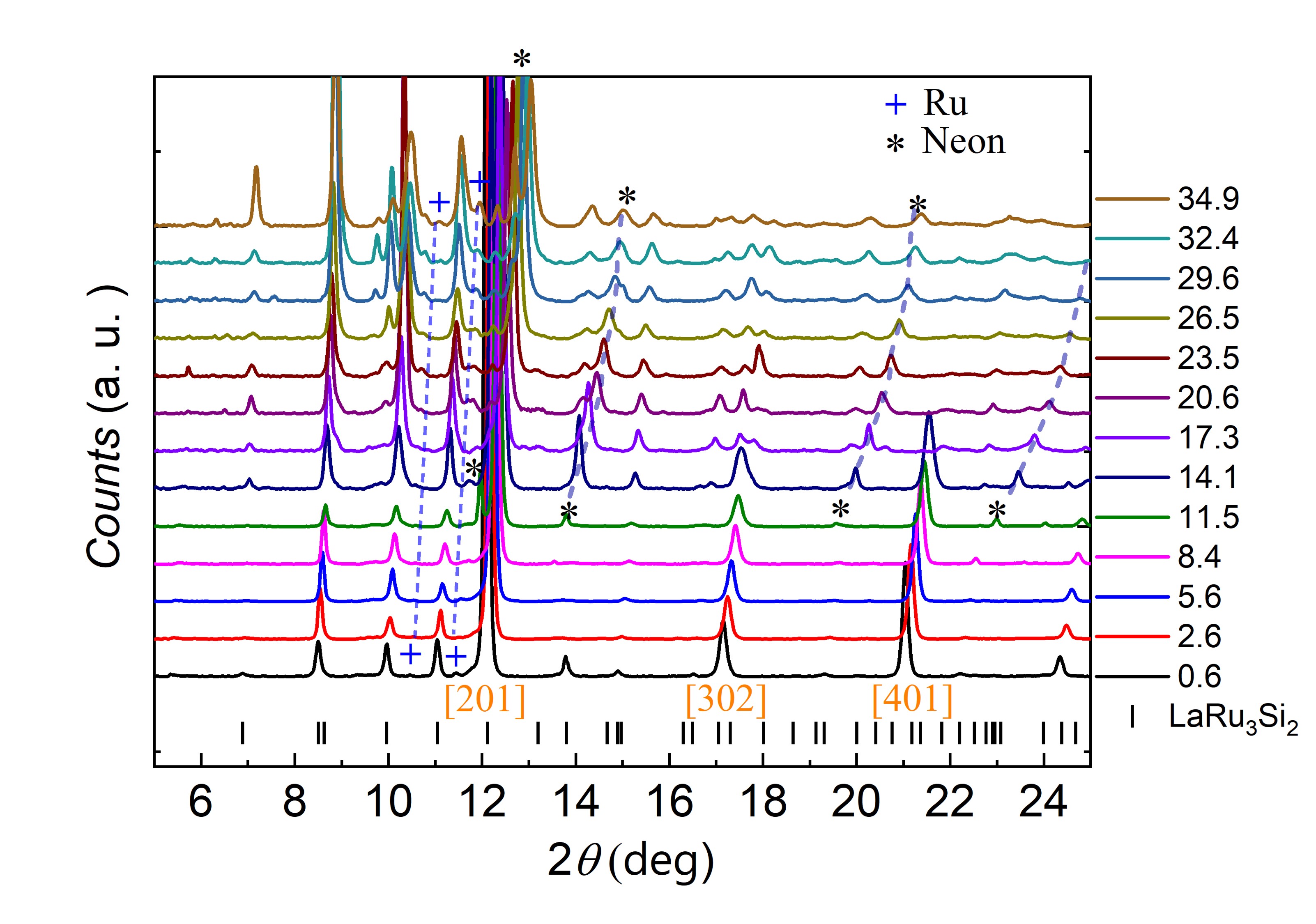}
\caption{\label{fig:HP_PXRD} Room-temperature PXRD patterns of LaRu$_3$Si$_2$ at various pressures up to 34.9 GPa. [201], [302], [401] peaks of LaRu$_3$Si$_2$ are labeled by orange brackets. Peak positions of Ru and neon are labeled by cross and star. Blue and black dashed lines show the evolution of Ru and neon diffraction peaks, respectively.}
\end{figure}

\begin{figure}[h]
\includegraphics[width=8cm]{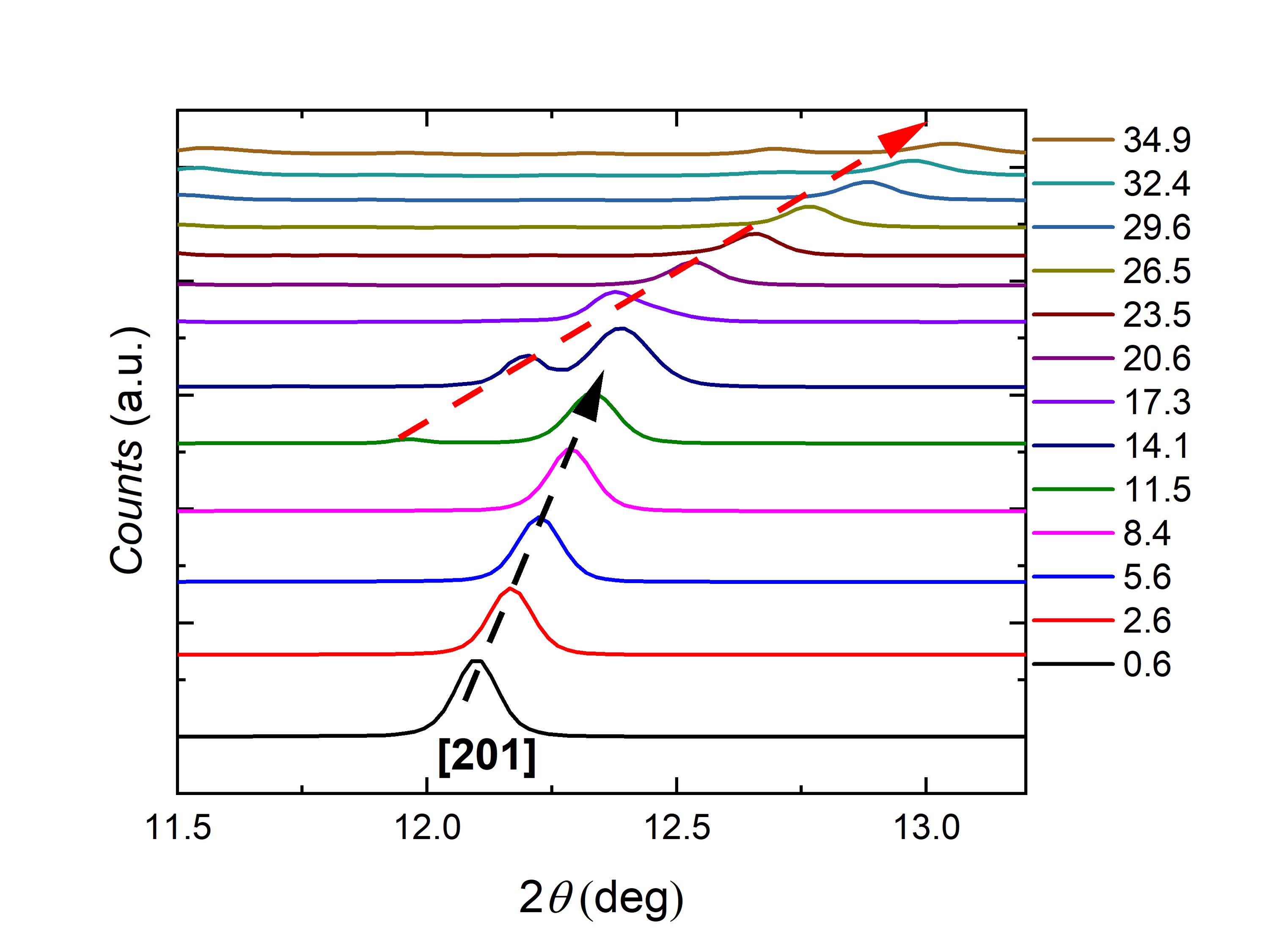}
\caption{\label{fig:HP_PXRD_201peak} PXRD pattern of [201] peak under pressure, black dashed line indicates the evolution of original [201] peak, and red dashed line shows the evolution of neon satellite peak.}
\end{figure}

Figure \ref{fig:HP_PXRD} shows room-temperature diffraction data collected across pressures up to $\sim$35 GPa. Whereas the hexagonal structural model perfectly describes the data up to $\sim$14 GPa, we note the following: (i) A sudden reduction in the intensities of specific peaks starting at $\sim$14 GPa, such as [201], [302], and [401], suggests either a subtle distortion within the hexagonal phase leading to a loss of long-range ordering or a preferred orientation developing under pressure without a structural transition, (ii) Distinct new diffraction peaks emerge gradually with increasing pressure from $\sim$23.5 GPa up to $\sim$34.9 GPa, as depicted in Fig.\ref{fig:HP_PXRD}, which suggests a second-order structural transition to a lower-symmetry phase The diffraction pattern could be assigned to an orthorhombic space group, \textit{Pmmm}; one possible solution of indexing \textit{Pmmm} peaks can be found in Appendix C, Fig.\ref{fig:Pmmm}.
(iii) Diffraction peaks from solidified neon, marked with black stars in Fig. \ref{fig:HP_PXRD}, appear at pressures above 11.5 GPa. As shown in Fig. \ref{fig:HP_PXRD_201peak}, one neon peak manifests as a satellite peak near the [201] peak. With increasing pressure, the [201] peak diminishes while the neon peak becomes dominant and shifts rapidly. This behavior raises the possibility that the subtle structural changes observed in LaRu$_3$Si$_2$ are influenced by non-hydrostatic conditions arising from neon solidification. It should be noted that whereas neon was used as a PTM in PXRD measurements, resistance was measured using the Nujol PTM, so solidification of the PTM cannot be the reason for the features observed in resistance and $T_c$  measurements in the 10 – 15 GPa pressure range. This observation potentially  gives some support to interpretation of PXRD data as an indication of a subtle structural phase transition at $\sim$14 GPa.

\section{\label{sec:level1}Discussion and Conclusion}

\subsection{Discussion}

\begin{figure}[h]
\includegraphics[width=8cm]{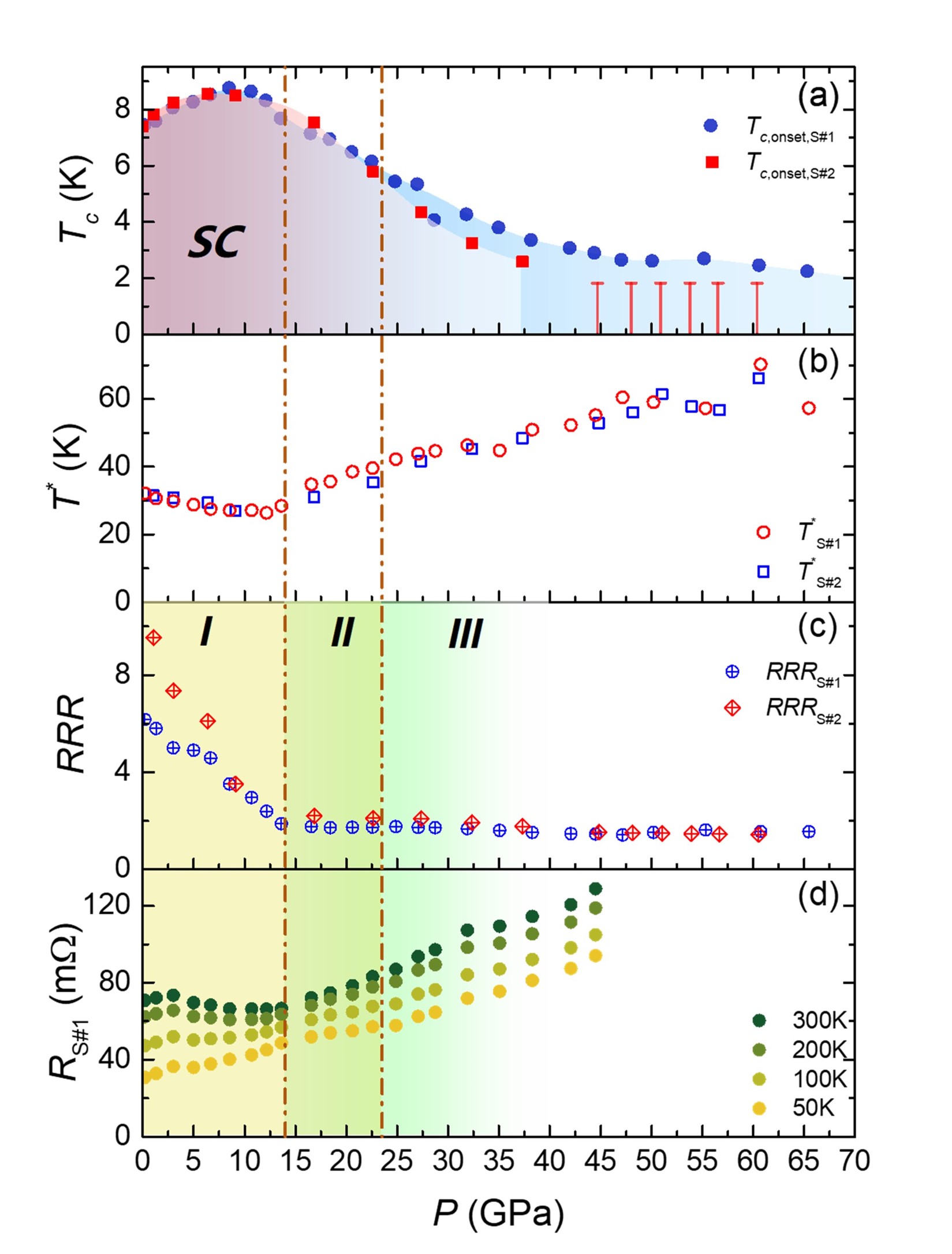}
\caption{\label{fig:TC_varP} Pressure dependence of (a) $T_c$, (b) $T^*$, (c) $RRR$ for both samples, along with (d) resistance at fixed temperature (300 K, 200 K, 100 K, 50 K) of sample S\#1. The superconducting dome is depicted by the blue shaded region (sample S\#1) and the red shaded region (sample S\#2) in (a); red vertical lines indicate the lowest temperature of our measurement (1.8 K). Brown vertical dashed lines separate three possible room temperature structure phases (\MakeUppercase{\romannumeral 1}, \MakeUppercase{\romannumeral 2}, \MakeUppercase{\romannumeral 3}).}
\end{figure}

Based on high-pressure transport and PXRD measurements on LaRu$_3$Si$_2$ samples, we constructed a temperature-pressure phase diagram in Fig. \ref{fig:TC_varP} (a). In addition, we plot $T^*(P)$, $RRR(P)$, and isothermal $R(P)$ in Fig. \ref{fig:TC_varP} (b-d) for comparison, with the shaded area denoting three possible room-temperature phases. As shown in Fig. \ref{fig:TC_varP} (a), $T_{c,\text{onset}}$ first increases with applying pressure, and reaches a maximum of $\sim$8.7K at $\sim$8.5 GPa. Then with further pressure, $T_{c,\text{onset}}$ drops resulting in a superconducting dome. This dome is consistent for both sample S\#1 and sample S\#2 until 35 GPa, at which point $T_{c,\text{onset}}$ of S\#1 starts to reach a flat plateau, whereas for S\#2, no superconducting transition was observed down to 1.8 K.

A similar superconducting dome was also reported in CeRu$_3$Si$_2$ \cite{CeRu3Si2_HP}, a mixed valent kagome superconductor with $T_c\simeq 1 \text{K}$ in the same structural family of LaRu$_3$Si$_2$. In the case of CeRu$_3$Si$_2$, the $T_c(P)$ behavior was explained on the basis of increased hybridization reducing the pair-breaking effect of Ce ions.Such a rationalization for $T_c$ enhancement cannot be applied to LaRu$_3$Si$_2$.

The evolution of $T^*$ is shown in Fig. \ref{fig:TC_varP} (b). $T^*$ first manifests a slow decrease from 30 K at ambient pressure to 25 K at $\sim$10 GPa, then starts increasing to 70 K, with no sign of limitation up to 65 GPa. Though $T_c$ differs slightly between samples S\#1 and S\#2 at higher pressure, $T^*$ remains consistent for the whole pressure range measured. 

Notably, distinct slope changes in $T^*$ and \textit{RRR} shown in Fig. \ref{fig:TC_varP} (b),(c) are observed around $\sim$14 GPa, aligning closely with the onset of a subtle structural transition (14.1 $<$ $P$ $<$ 17.3 GPa) at room temperature. This suggests a significant alteration in the electronic structure near the Fermi level. The progressive suppression of superconducting $T_c$, above 10 - 15 GPa, may result from a reduced density of states at the Fermi surface, as indicated by the pressure-induced increase in resistance across various temperatures. Alternatively, this suppression could be linked to an evolution of specific phonon modes in this lower symmetry structure under pressure. Further Raman spectroscopy studies under pressure may provide more insights into these mechanisms.

No systematic change of $T_c$, $T^*$, \textit{RRR}, and $R$(300 - 50 K) is observed in correspondence with the room-temperature structural transition from phase \MakeUppercase{\romannumeral 2} to \MakeUppercase{\romannumeral 3} near $\sim$23.5 GPa (shown in Appendix A, Fig. \ref{fig:S1_S2_varP}, for both S\#1 and S\#2). It is worth noting that room-temperature structure transition might affect low-temperature properties at different pressure, suggesting the possibility that $T_c$, $T^*$, \textit{RRR}, and $R$(300 - 50 K) exhibit distinct behavior outside this pressure range. 

Interestingly, another kagome superconductor family, \textit{A}V$_3$Sb$_5$ (\textit{A} = K, Rb, and Cs), exhibits two superconducting domes under pressure \cite{Zhu_2Dome_135_Rb_K,zhang_135_2dome}. The first dome is widely thought to be attributed to the suppression of CDW instability, whereas the origin of the second dome is still not well understood. In CsV$_3$Sb$_5$, the hexagonal structure remains stable under pressure, and the second dome has been associated with a Lifshitz transition \cite{zhang_135_2dome}. In contrast, KV$_3$Sb$_5$ and RbV$_3$Sb$_5$ display structural transitions, hexagonal-to-monoclinic and hexagonal to monoclinic to orthorhombic, respectively, which are believed to underpin the second dome \cite{Du_135_quasihydro}. Our findings in this work, on the other hand, provide valuable comparative insights and contribute to a deeper understanding of the exotic superconducting pairing mechanisms in kagome systems.

\subsection{Conclusion}
In conclusion, we have observed the superconducting dome of LaRu$_3$Si$_2$ along with two possible structural phase transitions at room temperature under pressure. Superconducting $T_c$ gets first enhanced to $\sim$8.7 K at $\sim$8.5 GPa and then suppressed with the further application of hydrostatic pressure. At almost the same pressure of $T_c$ suppression, the room temperature structure changes from the original structure (\textit{P6/mmm}) to a slightly distorted hexagonal structure, after which it maybe turn to a structure with lower symmetry, possibly orthorhombic, where $T_c$ suppression slows down. 

In addition, whereas the nature of $T^*$ is ambiguous, the possible correlation between $T^*$, superconductivity and structure demands investigations of temperature-dependent x-ray diffraction and calculations of electronic structure under pressure, to get further insight into this kagome metal system.

\textit{Note added.} As we submitted our work a posting on arXiv appeared \cite{ma2024domeshapedsuperconductingphasediagram} that presented a quantitatively very similar \textit{T-P} phase diagram.

\begin{acknowledgments}
The authors acknowledge Juan Schmidt and Atreyee Das for useful discussions. Work at Ames National Laboratory is supported by the U.S. Department of Energy, Office of Basic Energy Science, Division of Materials Sciences and Engineering. Ames National Laboratory is operated for the U.S. Department of Energy by Iowa State University under Contract No. DE-AC02-07CH11358. E.C.T. and W.B. acknowledge support from National Science Foundation (NSF) CAREER Award No. DMR-2045760. Portions of this work were performed at GeoSoilEnviroCARS (The University of Chicago, Sector 13), Advanced Photon Source, Argonne National Laboratory. GeoSoilEnviroCARS is supported by the National Science Foundation – Earth Sciences via SEES: Synchrotron Earth and Environmental Science (EAR –2223273). This research used resources of the Advanced Photon Source, a U.S. Department of Energy (DOE) Office of Science User Facility operated for the DOE Office of Science by Argonne National Laboratory under Contract No. DE-AC02-06CH11357.

ADDED in proof: As we submitted our work a posting on arXiv appeared [ref xx] that presented a quantitatively very similar T-P phase diagram \cite{ma2024domeshapedsuperconductingphasediagram}.
\end{acknowledgments}

\appendix

\section{High pressure transport measurement on sample S\#1 and S\#2}

\begin{figure}[h]
\includegraphics[width=8cm]{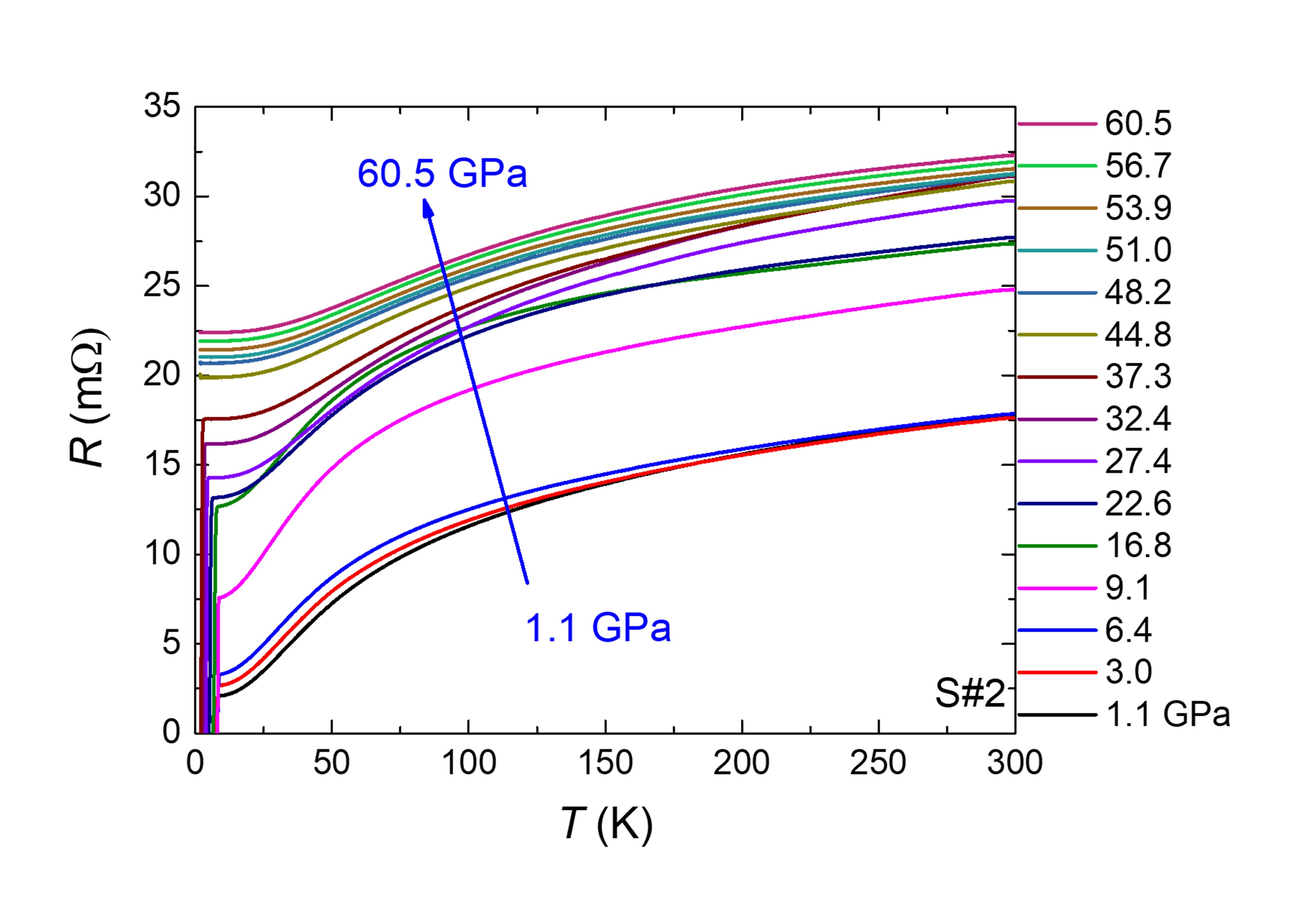}
\caption{\label{fig:S2_full} Temperature dependence of electrical resistance of sample S\#2 at various pressures.}
\end{figure}

High-pressure electrical transport measurements on samples S\#1 and S\#2 were performed using a diamond anvil cell (DAC) with the same pair of diamonds. Sample S\#1 and S\#2 were polished from two bulk samples in the same batch, while S\#2 was measured five months later than S\#1. The maximum pressure reached for sample S\#1 was 65.5 GPa and for sample S\#2 was 60.5 GPa.

\begin{figure}[h]
\includegraphics[width=8cm]{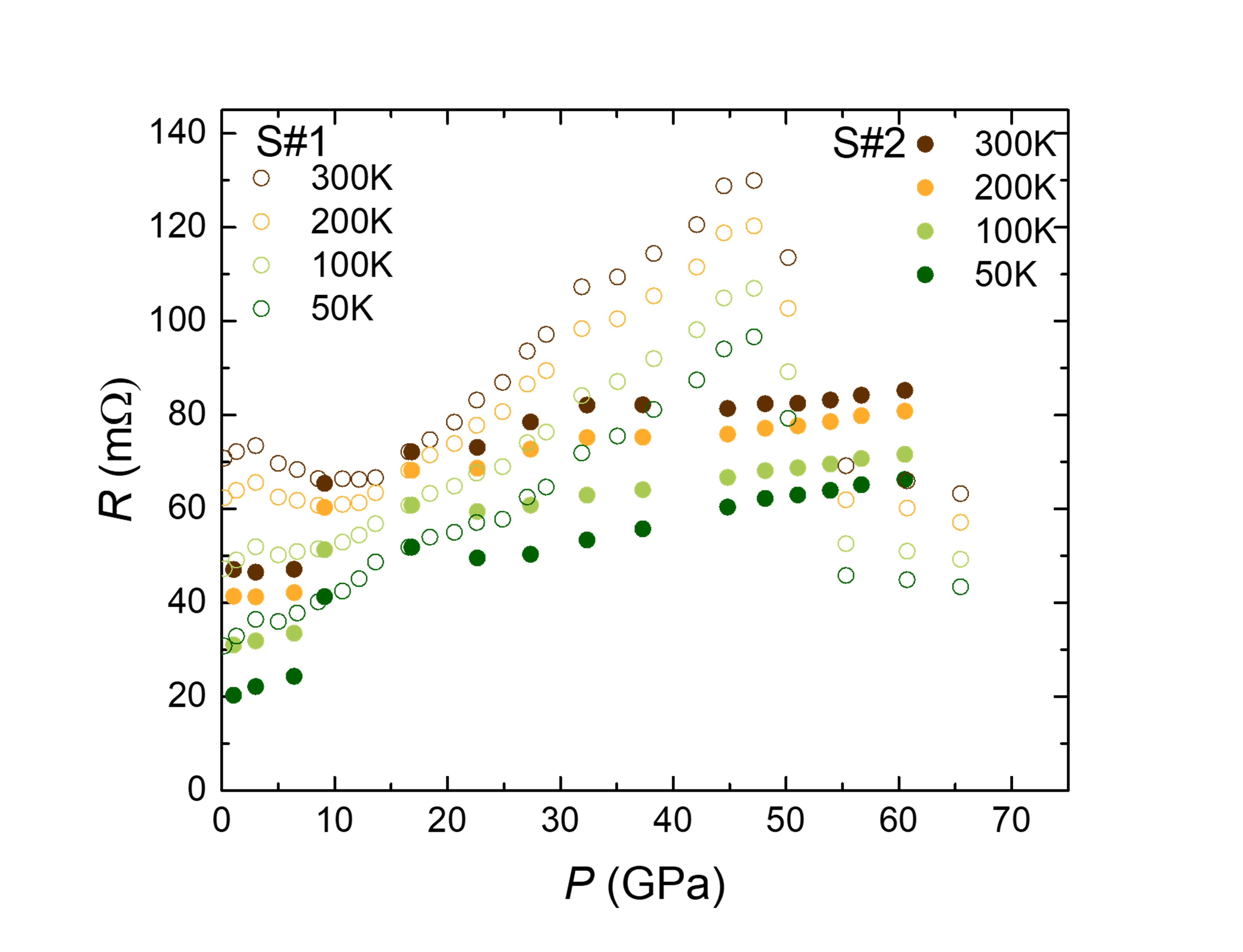}
\caption{\label{fig:S1_S2_varP} Resistances of S\#1 and normalized resistance (at 16.8 GPa) of S\#2 at various temperatures (300 K, 200 K, 100 K, 50 K) as function of pressure.}
\end{figure}

Figure \ref{fig:S2_full} shows the pressure dependence electrical resistance of sample S\#2 from 0 to 300 K at various pressures. There is a jump from the overall R(T) of 6.4 GPa to 9.1 GPa, which we supposed was due to the poor sample contact that was fixed at 9.1 GPa. Sample S\#2 shows a weaker pressure dependence compared to S\#1, as show in Fig. \ref{fig:S1_S2_varP}. To compare the two samples, the resistance of S\#2 in Fig. \ref{fig:S1_S2_varP} has been normalized to that of S\#1 at $\sim$17 GPa at each temperature individually. In the $\sim$10 GPa to $\sim$50 GPa region they are both following an increasing trend. However, the sudden drop of S\#1 at 50 GPa was not repeated for S\#2; instead S\#2 keeps following an increasing trend. Although it is difficult to rule out the possible sample’s variance and inhomogeneity, such an anomaly in S\#1 of sudden pressure drop is more likely due to experimental artifacts coming from electrical contacts, which has less impact on the intrinsic SC transition temperature. Thus, $T_c$ for S\#1 at pressure $>$ 50 GPa is still taken for drawing the phase diagram in Fig. \ref{fig:TC_varP} (a).   Such variance between two samples does not affect our main conclusion.

\section{Linear and WHH fit for ambient and high pressure transport measurement}

\begingroup
\squeezetable
\begin{table}[b]
\caption{\label{tab:Hc2_table}
Summarized upper critical fields under different fitting model with $T_{c,\text{onset}}$ at various pressures.}
\begin{ruledtabular}
\begin{tabular}{ccccccc}
 $P$ (GPa) & ambient &1.3 &3.0 &16.6&42.1 &50.2 \\
\hline
$H_{c2,\text{LF}}(0)$ (kOe) \footnotemark[1] & 71.2& 62.2 & 69.5 & 61.5 & 29.6 & 30.6 \\
$H_{c2,\text{WHH}}(0)$(kOe) & 45.1 & 47.0 & 50.5 & 45.9 & 19.2 & 17.4 \\
$H_{P}$(kOe) & 130.1 & 138.2 & 148.5 & 134.1 & 55.6& 47.6 \\
-$\cfrac{dH_{c2}}{dT}(T = T_c)$(kOe/K) & 9.2 & 9.0 & 9.1 & 10.6 & 10.9 & 11.4 \\
\end{tabular}
\end{ruledtabular}
\footnotetext[1]{LF stands for linear fit}
\end{table}
\endgroup

The Werthamer–Helfand–Hohenberg (WHH) model provides a prediction of the upper critical field $H_{c2}(T)$ in type-II superconductors, taking account of the effect of both orbital and paramagnetic effects. In the dirty limit, where electron mean free path $l$ is much smaller than superconducting coherence length $\xi$, $H_{c2}(T)$ follow:

\begin{equation}
    \begin{aligned}
    &\text{ln}\frac{1}{t}=\left(\frac{1}{2}+\frac{i \lambda_{SO}}{4 \gamma}\right) \psi\left(\frac{1}{2}+\frac{\bar{h}+\frac{1}{2} \lambda_{SO}+i \gamma}{2 t}\right) \\
    &+\left(\frac{1}{2}-\frac{i \lambda_{SO}}{4 \gamma}\right) \psi\left(\frac{1}{2}+\frac{\bar{h}+\frac{1}{2} \lambda_{SO}-i \gamma}{2 t}\right)+c.c.
    \end{aligned}
\end{equation}
where $t = T/T_c$ is the normalized temperature, $\bar{h} = 4/\pi^2\cdot dH_{c2}/(dH_{c2}/dt)_{t=1}$ is the reduced magnetic field, and $\gamma =\left[(\alpha \bar{h})^2-(\frac{1}{2}\lambda_{SO})^2\right]^{1/2}$, Maki parameter $\alpha$ and spin-orbit coupling parameter $\lambda_{SO}$ are fitting parameters.

\begin{figure}[h]
\includegraphics[width=8cm]{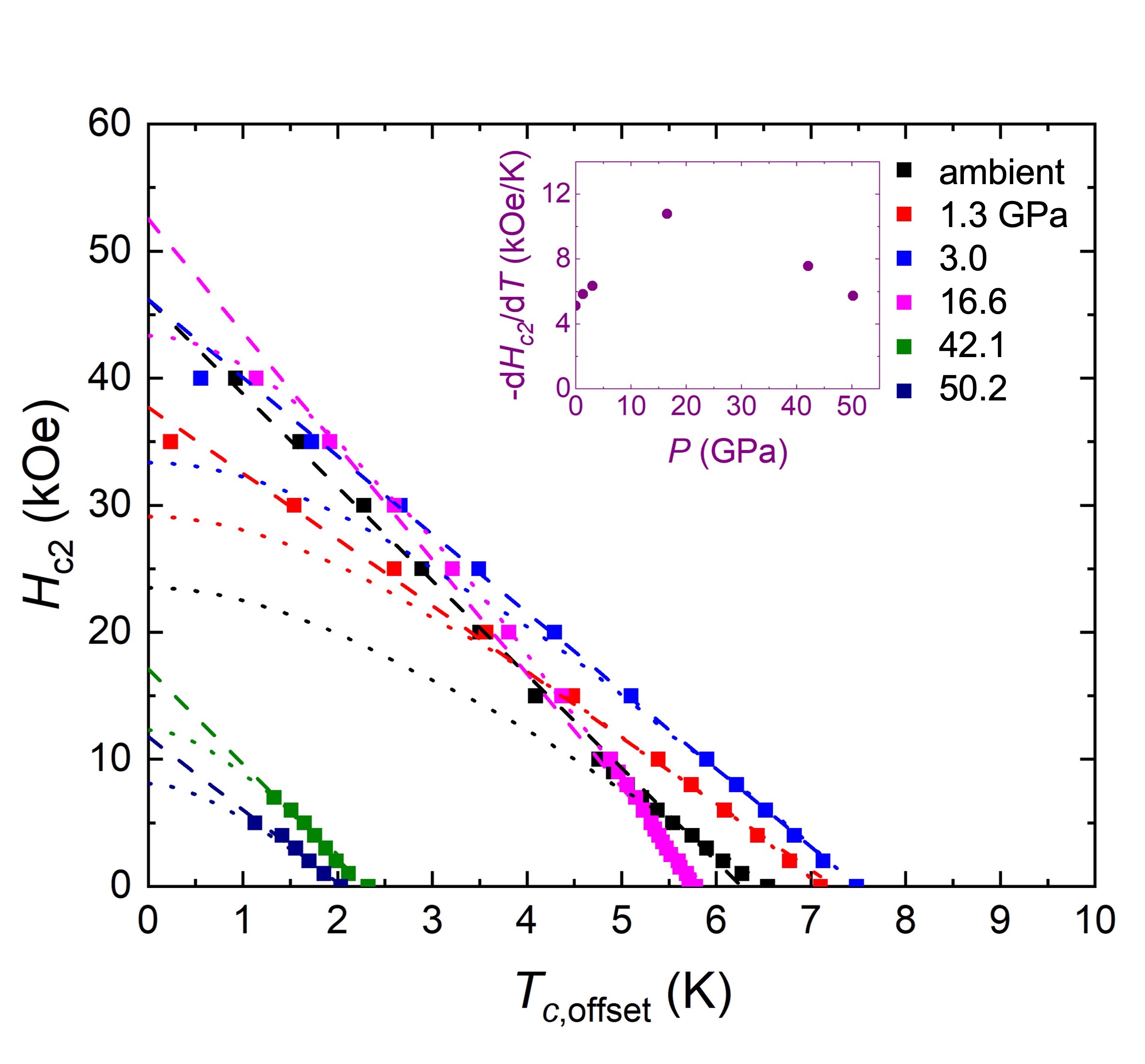}
\caption{\label{fig:Hc2_varP_offset} Upper critical field $H_{c2}$ dependence of $T_{c,\text{offset}}$ obtained from transport measurements at several pressures, inset showing the temperature derivative of $H_{c2}$ at $T_{c,\text{offset}}$. Dashed and dotted lines are the fitted $T$ linear lines and WHH fitted lines for each pressure, respectively. %dotted lines are fitted WHH model lines.
}
\end{figure}

Tables \ref{tab:Hc2_table} and \ref{tab:Hc2_table_2} give the summary of $H_{c2}(T)$ at various pressures for both linear fit, WHH model, -$\cfrac{dH_{c2}}{dT}(T = T_c)$, as well as the BCS weak-coupling Pauli paramagnetic limit $H_P=18.3T_c$ (kOe) \cite{Tinkham_SC}, fitted with $T_{c,\text{onset}}$ and $T_{c,\text{offset}}$ individually. 

Figure \ref{fig:Hc2_varP_offset} shows the plot of fitting results of $H_{c2}$ with the $T_{c,\text{offset}}$ criterion. It is worth noting that generally the $H_{c2}(T)$ dependence shows the same linear feature as fitted with $T_{c,\text{onset}}$, deviating from the WHH model. However the slope -$\cfrac{dH_{c2}}{dT}(T = T_c)$ got almost doubled at 16.6 GPa. Such significant difference remains unclear at this point, and it could arise from the sample's inhomogeneities and multi step transition (as shown in FIG. \ref{fig:RT_varH_varP} (d)).

\begingroup
\squeezetable
\begin{table}
\caption{\label{tab:Hc2_table_2}
Summarized upper critical fields under different fitting models with $T_{c,\text{offset}}$ at various pressures.}
\begin{ruledtabular}
\begin{tabular}{ccccccc}
 $P$ (GPa) & ambient &1.3 &3.0 &16.6&42.1 &50.2 \\
\hline
$H_{c2,\text{LF}}(0)$ (kOe) & 46.2&  37.7& 46.2 & 52.5 & 17.1 & 11.8  \\
$H_{c2,\text{WHH}}(0)$(kOe) &  23.5 & 29.1 & 33.3 & 43.3 & 12.3 & 8.1   \\
$H_{P}$(kOe) & 119.7 & 129.9 & 136.9 & 105.8 & 42.6 &  37.2 \\
-$\cfrac{dH_{c2}}{dT}(T = T_c)$(kOe/K) & 5.1 & 5.9 & 6.4 & 10.8 & 7.6 & 5.7   \\
\end{tabular}
\end{ruledtabular}
% \footnotetext[1]{LF stands for linear fit}
\end{table}
\endgroup

\section{Powder X-ray diffraction under high pressure refinement and indexing \textit{Pmmm} peaks}

\begin{figure}[h]
\includegraphics[width=8.5cm]{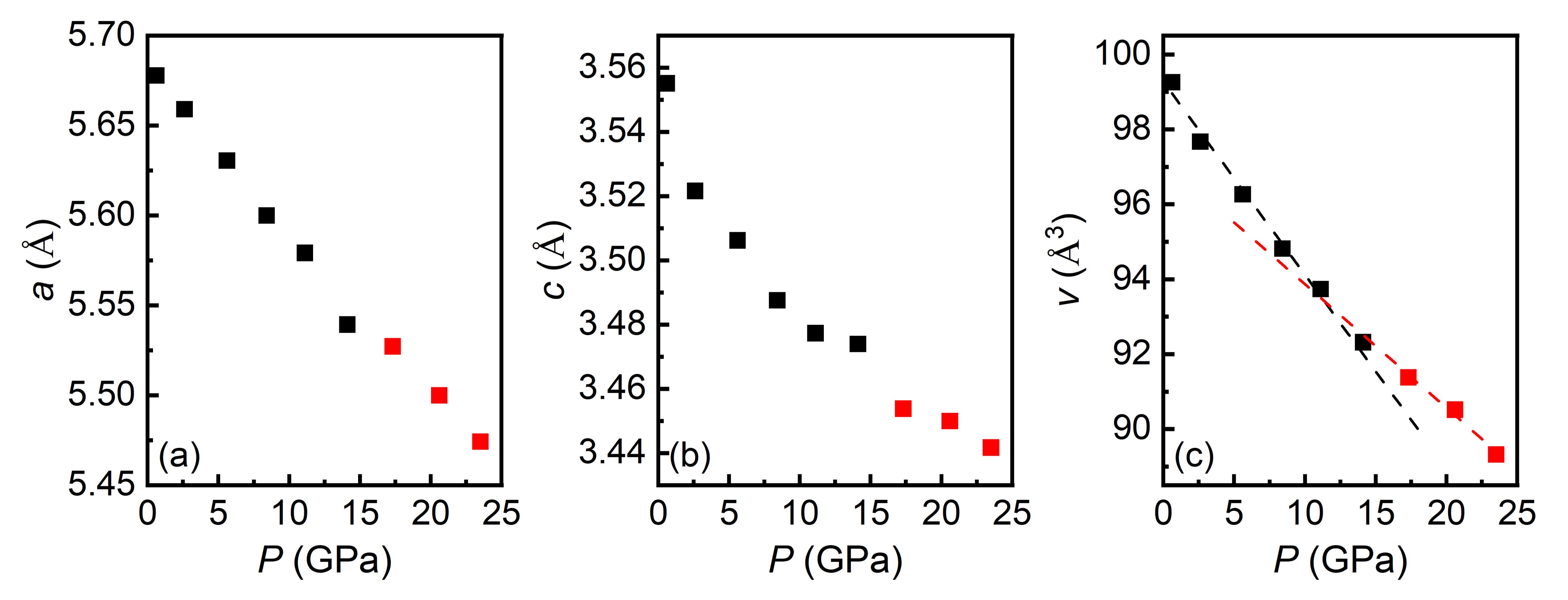}
\caption{\label{fig:Lattice_varP} Lattice parameters \textit{a}, \textit{c} and (c) unit cell volume \textit{v} evolution with pressure, under the assumption that crystal symmetry remains as \textit{P6/mmm}. The dashed lines in panel (c) are guides to the eye.}
\end{figure}

\begin{figure}[h]
\includegraphics[width=8.5cm]{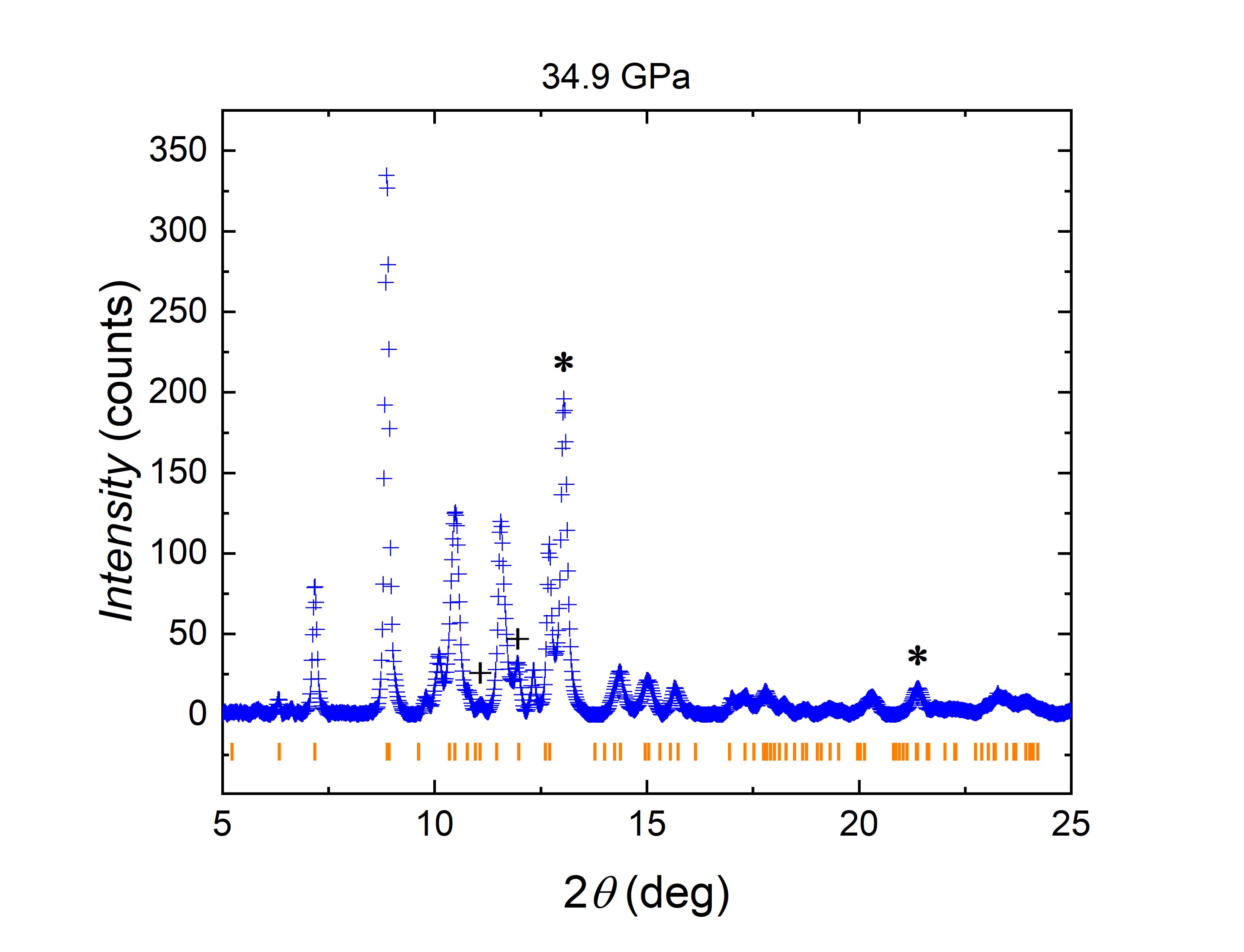}
\caption{\label{fig:Pmmm} Indexed \textit{Pmmm} peaks of PXRD patterns at 34.9 GPa. Orange line marks stand for indexed peaks of \textit{Pmmm} structure, with lattice parameter $a=2.743$ \r A, $b=4.679$ \r A, $c=6.824$ \r A. Peaks marked with cross and star are Ru and neon peaks, respectively.}
\end{figure}

Room-temperature high pressure powder x-ray diffraction (PXRD) experiments were conducted using a grounded sample from the same batch as high-pressure electrical transport. Neon was used as the pressure-transmitting medium (PTM).

Figure \ref{fig:Lattice_varP} (a-c) shows the pressure dependence of lattice parameters \textit{a} and \textit{c}, along with the unit cell volume $v$, assuming the crystal symmetry remains unchanged up to 23.5 GPa. In FIG. \ref{fig:Lattice_varP} (a) a subtle discontinuity can be found from 14.1 GPa to 17.3 GPa for \textit{a}. In FIG. \ref{fig:Lattice_varP} (b) there is a clearer jump of \textit{c} at same pressure. Meanwhile the unit cell volume \textit{v} shows a change of slope at the transition. These changes are observed in the pressure range where the PTM, neon, solidifies. Further work is required to sort out possible effects of non-hydrostaticity in these measurements.

Figure \ref{fig:Pmmm} shows the indexed peaks for one possible \textit{Pmmm} structure solution at 34.9 GPa, the unit cell parameters identified as  
$a=2.743$ \r A, $b=4.679$ \r A, and $c=6.824$ \r A. Note that the structure of Ru has been reported to be stable up to 150 GPa \cite{anzellini19_pureRu}, thus we can exclude the possibility that new peaks come from Ru impurity.

\newpage
\nocite{*}

\bibliography{apssamp}% Produces the bibliography via BibTeX.

\end{document}